\documentclass[fleqn,usenatbib]{mnras}

\usepackage[T1]{fontenc}

\DeclareRobustCommand{\VAN}[3]{#2}
\let\VANthebibliography\thebibliography
\def\thebibliography{\DeclareRobustCommand{\VAN}[3]{##3}\VANthebibliography}

\usepackage{graphicx}	
\usepackage{amsmath}	
\usepackage{amssymb}	
\usepackage{multicol}   
\usepackage{bm}		    
\usepackage{pdflscape}	
\usepackage{multirow}
\usepackage{booktabs}
\usepackage{array}
\usepackage{float}
\usepackage{xspace} 
\usepackage{makecell}

\newcommand{\dd}{\partial}

\newcommand{\tr}{\ \mathrm{tr}}
\newcommand{\inv}{^{\raisebox{.2ex}{$\scriptscriptstyle-1$}}}
\newcommand{\mat}[1]{\mathbf{#1}}
\newcommand*\diff{\mathop{}\!\mathrm{d}}

\newcommand*{\eg}{\emph{e.g.}\@\xspace}
\newcommand*{\ie}{\emph{i.e.}\@\xspace}

\newcommand{\Planck}{\emph{Planck}\xspace}
\newcommand{\WMAP}{\emph{WMAP}\xspace}
\newcommand{\commander}{\texttt{COMMANDER}\xspace}
\newcommand{\SRone}{\texttt{SRoll1}\xspace}
\newcommand{\SRtwo}{\texttt{SRoll2}\xspace}
\newcommand{\simbal}{\textsc{SimBaL}\xspace}
\newcommand{\simlow}{\texttt{C-SimLow}\xspace}
\newcommand{\glass}{\textsc{glass}\xspace}
\newcommand{\momento}{\texttt{momento}\xspace}
\newcommand{\delfi}{\textsc{delfi}\xspace}
\newcommand{\pydelfi}{\texttt{pydelfi}\xspace}
\newcommand{\camspec}{\texttt{CamSpec}\xspace}

\newcommand{\tauEEplanckofficial}{$\tau_{EE}^{\SRone}=0.0506 \pm 0.0086$\xspace}
\newcommand{\tauEEsrollofficial}{$\tau_{EE}^{\SRtwo}=0.0566^{\ +0.0053}_{\ -0.0062}$\xspace}

\def\ltsima{$\; \buildrel < \over \sim \;$}
\def\gtsima{$\; \buildrel > \over \sim \;$}
\def\simlt{\lower.5ex\hbox{\ltsima}}
\def\simgt{\lower.5ex\hbox{\gtsima}}

\providecommand{\sorthelp}[1]{}

\newcolumntype{N}{>{\centering\arraybackslash}m{.5in}}
\newcolumntype{G}{>{\centering\arraybackslash}m{2in}}

\defcitealias{2016A&A...596A.107P}{PSRoll1}
\defcitealias{2019A&A...629A..38D}{PSRoll2}
\defcitealias{Aghanim:2018eyx}{PCP18}
\defcitealias{2020A&A...641A...5P}{PLP19}
\defcitealias{EG19}{EG19}

\usepackage{newtxtext,newtxmath}

\title[Inference of $\tau$ using Planck cross-spectra]{Inference of the optical depth to reionization from low multipole temperature and polarisation \Planck data}

\author[de Belsunce et al.]{
Roger de Belsunce,$^1$\thanks{E-mail: \href{mailto:rmvd2@cam.ac.uk}{rmvd2@cam.ac.uk}}
Steven Gratton,$^{1}$
William Coulton,$^{2}$
and George Efstathiou$^1$
\\
$^{1}$Kavli Institute for Cosmology \& Institute of Astronomy, University of Cambridge, Madingley Road, Cambridge CB3 OHA, United Kingdom\\
$^{2}$Kavli Institute for the Physics and Mathematics of the Universe (WPI), The University of Tokyo Institutes for Advanced Study (UTIAS),\\
$\thinspace \ $The University of Tokyo, Kashiwa, Chiba 277-8583, Japan}

\date{Accepted XXX. Received YYY; in original form ZZZ}

\pubyear{2021}

\begin{document}
\label{firstpage}
\pagerange{\pageref{firstpage}--\pageref{lastpage}}
\maketitle

\begin{abstract}
This paper explores methods for constructing low multipole temperature and polarisation likelihoods from maps of the  cosmic microwave background anisotropies that have complex noise properties and partial sky coverage. We use \Planck 2018 High Frequency Instrument (HFI) and updated \SRtwo temperature and polarisation maps to test our methods. We present three likelihood approximations based on quadratic cross spectrum estimators: (i) a variant of the simulation-based likelihood (\simbal) techniques used in the \Planck legacy papers to produce a low multipole $EE$ likelihood; (ii) a semi-analytical likelihood approximation (\momento) based on the principle of maximum entropy; (iii) a density-estimation `likelihood-free' scheme (\delfi). Approaches (ii) and (iii) can be generalised to produce low multipole joint temperature-polarisation ($TTTEEE$) likelihoods. We present extensive tests of these methods on simulations with realistic correlated noise. We then analyse the \Planck data and confirm the robustness of our method and likelihoods on multiple inter- and intra-frequency detector set combinations of \SRtwo maps. The three likelihood techniques give consistent results and support a low value of the optical depth to reoinization, $\tau$,  from the HFI. Our best estimate of $\tau$ comes from combining the low multipole \SRtwo \momento ($TTTEEE$) likelihood with the \camspec high multipole likelihood and is $\tau =  0.0627^{+0.0050}_{-0.0058}$. This is consistent with the \SRtwo team's determination of $\tau$, though slightly higher by $\sim 0.5\sigma$, mainly because of our joint treatment of temperature and polarisation.
\end{abstract}


\section{Introduction}

Over the last decade, observations of the cosmic microwave
  background (CMB) \citep{2013ApJS..208...19H, Aghanim:2018eyx,
    Henning:2018, Aiola:2020}, together with measurements of the baryon
  acoustic oscillation scale from large galaxy surveys
  \citep{Gil-Marin:2020, Bautista:2020} and many other cosmological
  observations have transformed cosmology into a high precision
  science.  In cosmological data analysis, an accurate representation
  of the likelihood, as well as the ability to model systematics,
  are crucial in order to make reliable inferences from
  data. Exact likelihoods are often either
  unknown or computationally expensive to compute.  In addition,
  systematics in the data may bias the results if they cannot be
  modelled with fidelity and included in the likelihood. 

These issues are of particular importance for the measurement of
the optical depth to reionization $\tau$
from \Planck temperature and polarisation CMB maps.
Heuristic likelihood models for CMB data on a cut sky  with idealised noise
properties  have been proposed by 
\emph{e.g.} \citet{2008PhRvD..77j3013H,  2015MNRAS.453.3174M}. 
However, the accuracy of these models is
difficult to quantify, especially for cross-correlations of \Planck polarisation maps
which  have complex noise correlations and systematics. For these
reasons, the Planck collaboration adopted a simulation-based approach to
construct a low multipole polarisation likelihood from the HFI \Planck maps \citep[][hereafter \citetalias{2016A&A...596A.107P}]{2016A&A...596A.107P}.

In this paper, we apply three likelihood approximations to measure the optical depth to reionization
from \Planck HFI maps. All three methods use 
Bayesian statistics to make inferences about models from data. 
 Bayes' theorem can be used to infer the posterior density $\mathcal{P}(\bm{\theta} \vert \mathbf{d}_0, \mathcal{M})$ 
of a set of  parameters  $\bm{\theta}$ describing a  model $\mathcal{M}$ from a realisation of data $\mathbf{d}_0$:
\begin{equation}
    \label{eq:Bayes_theorem}
   \mathcal{P}(\theta|\mathbf{d}_0, \mathcal{M}) = \frac{\mathcal{P}(\mathbf{d}_0|\theta, \mathcal{M}) \mathcal{P} (\theta| \mathcal{M})}{\mathcal{P}(\mathbf{d}_0| \mathcal{M})} \Leftrightarrow \mathcal{P}_{\mathbf{d}} =  \frac{\mathcal{L}_{\mathbf{d}}\pi}{\mathcal{Z}_{\mathbf{d}}}\ , 
\end{equation}
where $\mathcal{P}_{\mathbf{d}}$ is the posterior,
$\mathcal{L}_{\mathbf{d}}$ the likelihood, $\pi$ the prior and
$\mathcal{Z}_{\mathbf{d}}$ the evidence. The subscript $\mathbf{d}$
denotes the dependence on the data set.  We compare the
simulation-based likelihood (\simbal) method of \citetalias{2016A&A...596A.107P}, which was used in the \Planck 2018 analysis of cosmological parameters \citep[][hereafter \citetalias{Aghanim:2018eyx}]{Aghanim:2018eyx},  with a flexible semi-analytic
likelihood approximation \citep[\glass; ][]{2017arXiv170808479G} and a density-estimation `likelihood-free' method\footnote{`Likelihood-free' (LF) methods are clearly not likelihood-free. What is meant is that the likelihood $\mathcal{L}$ is inferred by fitting to numerical simulations rather than being expressed as a simple functional form.} \citep[\delfi; ][]{2019MNRAS.488.4440A}. \glass can easily be adapted to produce a joint temperature-polarisation likelihood at low multipoles. However, this is nontrivial for \simbal and \delfi when trying to achieve near optimal results.

The optical depth to reionization provides a  measure of the time at which the intergalactic medium (IGM) was
reionized by photons produced by early generations of stars and galaxies. Following recombination at 
$z \sim 1000$, the  IGM remained almost neutral until reionization. Assuming  abrupt reionization at  $z_{\rm re}$,
the Thomson optical depth $\tau$ is
\begin{equation} \label{eq:tau_eq}
    \tau = \frac{2c\sigma_T (1-Y_{\rm p})}{m_{\rm p}}\frac{\Omega_{\rm b}}{\Omega_{\rm m}} \frac{H_0}{8\pi G} \left( \sqrt{\Omega_{\rm m} (1+z_{\rm re})^3 + \Omega_{\Lambda}}-1 \right) \ , 
\end{equation}
where $\sigma_T$ is the Thomson cross-section and we have assumed the
base $\Lambda$CDM model\footnote{As in the \Planck papers we refer to
  the six parameter $\Lambda$CDM model (spatially flat, power law scalar
  adiabatic fluctuations, cosmological constant) as the base
  $\Lambda$CDM model.} with a helium abundance by mass of $Y_{\rm P}$
(assuming that helium remains neutral).  The Gunn-Peterson test
\citep{1965ApJ...142.1633G, 2006AJ....132..117F} provides strong
astrophysical evidence that the intergalactic medium was highly
ionised by a redshift of $z = 6.5$. Using the \Planck 2018 base
$\Lambda$CDM parameters, Eq. \eqref{eq:tau_eq} yields a lower limit of
$\tau \approx 0.04$ for $z_{\rm re} =6.5$.

In this paper, we have measured  $\tau$ from two sets of \Planck maps: the \Planck 2018 legacy release \citep{LFI_data:2018, HFI_data:2018}, which for HFI is based on the \SRone map-making algorithm described in \citetalias{2016A&A...596A.107P},
and the improved map-making algorithm \SRtwo \citep[][hereafter \citetalias{2019A&A...629A..38D}]{2019A&A...629A..38D}.  The values of $\tau$ computed from the
low multipole $EE$ spectra alone\footnote{It is important to distinguish constraints on $\tau$ derived using a
low multipole likelihood alone, together with a constraint on the parameter combination $A_s \exp(-2\tau)$,
from full Monte Carlo Markov Chain (MCMC)  explorations that combine a low multipole likelihood with a high multipole likelihood.
For most of this paper, we will analyse low multipole likelihoods via single parameter scans through values of $\tau$,
and so we use Eqs. \eqref{eq:Tau1} and \eqref{eq:Tau2} as our references to previous work. Full MCMC parameter searches
combining our low multipole likelihoods with a high multipole $TTTEEE$ likelihood are deferred until Sec. \ref{sec:cosmo}.
For reference, assuming the base six parameter $\Lambda$CDM model, and adding the $\texttt{Plik}$ high multipole likelihood, the best fit
values of $\tau$ are higher than those of Eqs. \eqref{eq:Tau1} and \eqref{eq:Tau2} by about $0.5\sigma$ (see Eqs. \eqref{eq:tau_TTTEEE_cosmo1} and \eqref{eq:tau_TTTEEE_cosmo2}).}
from these maps are:
\begin{subequations}
\begin{align}
\qquad    &\text{\tauEEplanckofficial},  \ \quad {\text{ \citepalias{Aghanim:2018eyx}}}, \label{eq:Tau1} \\
\qquad    &\text{\tauEEsrollofficial}, \qquad {\text{ \citep{2019arXiv190809856P}}} . \label{eq:Tau2}
\end{align}
\end{subequations}

These estimates improve significantly on the result from  \WMAP \citep{2013ApJS..208...19H} of $\tau=0.089\pm0.014$. Measurements of $\tau$ using a pixel-based likelihood on Low Frequency Instrument (LFI) \Planck and \WMAP data have been presented in \citet{2017JCAP...02..041L} and \citet{2020A&A...644A..32N}, albeit with larger uncertainties than in Eqs. \eqref{eq:Tau1} and \eqref{eq:Tau2}.
Note that the estimates from \Planck are just above the Gunn-Peterson limit of $\tau \sim 0.04$ inferred from Eq. \eqref{eq:tau_eq}.
This implies that reionization occurred late, i.e. $z_{\rm re}$ cannot be much greater than about $6.5$ \citep[see for example][and references therein]{Kulkarni:2019}. 

Measuring $\tau$ using the CMB is challenging. At high multipoles, the CMB power spectra are damped by a factor 
$e^{-2\tau}$ leading to a degeneracy between the $\tau$ parameter and the amplitude of the initial cosmological scalar perturbations $A_s$ (which is partially broken by CMB lensing). Reionization also induces a  polarisation signal 
at super-horizon scales in the $EE$ power spectrum leading to a `reionization bump' at low multipoles ($\ell \simlt 20$) with an 
amplitude that scales  approximately as $\tau^2$. The $EE$ power spectrum at low multipoles
can therefore be used to constrain $\tau$, provided systematics can be kept under control.

An enormous effort has been made to improve the fidelity of the
  \Planck HFI polarisation maps. These improvements are presented in
  detail in \citetalias{2016A&A...596A.107P} and \citetalias{2019A&A...629A..38D}. In the analysis described in \citetalias{2019A&A...629A..38D},
   residual systematics at $100$ and $143$ GHz are reduced to
  levels below the notional detector noise levels at multipoles $\ell
  \simlt 10$,  as demonstrated by a number of null tests. \citetalias{2016A&A...596A.107P} and \citetalias{2019A&A...629A..38D}
  describe sets of end-to-end simulations of the two \texttt{SRoll} pipelines. In
  this paper, we use these simulations to characterise large-scale systematic modes 
  and correlated   noise at low multipoles.  We then
  generate a large number of realisations with realistic noise and
  systematics over a grid of $\tau$ values, which are used to train, or
  calibrate, two of the likelihood models. All of the likelihoods are based on a
  quadratic cross spectrum (QCS) estimator, which we use to measure
  the foreground cleaned cross-spectra at low
  multipoles ($2\leq\ell \leq29$) from the \SRone and \SRtwo
  temperature and polarisation maps\footnote{Most of our results are
based on the $100\times143$ full-mission cross-spectra, though we also
investigate $100 \times 100$ and $143 \times 143$  `detector set' cross-spectra.}.

This paper is organised as follows: in Sec.\ \ref{sec:CMB_QCS_estimator}
we review the QCS power spectrum estimator. In
Sec. \ref{sec:data}, we discuss map compression and foreground
cleaning procedures as well as residual systematics in the maps and their
contribution to the power spectrum. In Sec.\ \ref{sec:NCM} we derive the
pixel-pixel noise covariance matrices required for the QCS estimator
and the likelihood computations. In Sec.\ \ref{sec:Likelihood_sec} we
present the different likelihood methods used to measure $\tau$: the simulation-based
likelihood (\simlow) in Sec.\ \ref{sec:Likelihood_sim}, the likelihood
approximation scheme (\momento) in Sec.\ \ref{sec:Likelihood_GLASS} and
the likelihood-free approach (\pydelfi) in
Sec.\ \ref{sec:Likelihood_DELFI}. In Sec. \ref{sec:compare_likelihoods} we test our three likelihoods on simulations with realistic correlated noise. In
Sec.\ \ref{sec:posteriors} we analyse the \Planck data and
perform cross-checks to validate our results. 
Sec.\ \ref{sec:Conclusions} presents our conclusions.

\section{quadratic cross spectrum estimator} \label{sec:CMB_QCS_estimator}

To make inferences from large data sets such as CMB maps, data
  compression is often required to reduce the size of the data vector
  to a manageable level. Here we compress maps into summary
  statistics, namely the angular power spectra\footnote{Throughout the paper, we use the following notation: $\tilde{C}_{\ell}$ are the un-deconvolved and $\Hat{C}_{\ell}$ the deconvolved power spectra of the data. Theory spectra are denoted by $C_{\ell}$.  
  } $C^r_{\ell}$, for each mode ($r
  \equiv TT, TE, EE, BB \dots$).  Quadratic estimators can be
  constructed to measure the temperature and polarisation power
  spectra on an incomplete sky which have lower variance than
  traditional pseudo-$C_{\ell}$ (PCL) estimators
  \citep{Tegmark:2001zv, 2006MNRAS.370..343E} and are easily
  computable at low multipoles from low resolution maps\footnote{It is
    also possible to write down a pixel-based likelihood for low
    resolution maps, provided the signal and noise are Gaussian and
    the noise covariance matrix $N_{ij}$ is known accurately, see \citet{Page:2007, 2017JCAP...02..041L, 2020A&A...644A..32N}.}.

Ordering the $T$, $Q$ and $U$ pixel values as a data vector  ${\bf x}$,  one can write down a 
quadratic power spectrum estimate $\mat y^r_{\ell}$ \citep{PhysRevD.55.5895}:
\begin{equation}
 \mat y^r_{\ell} = x_i x_j \mat E^{r \ell}_{ij} \ , \label{ML1a}
\end{equation}
where  
\begin{equation}
 \mat E^{r \ell} = \frac{1}{2} \mat C\inv \frac{\partial \mat C}{\partial C^r_\ell} \mat C\inv \ , \label{ML1b}
\end{equation}
$\mat C$ is the covariance matrix of the data vector ${\bf x}$, 
\begin{equation} \label{eq:cov_QCS}
\mat C=\langle x_i x_j\rangle = 
\begin{pmatrix}
\mat C^{TT}&\mat C^{TQ}&\mat C^{TU}\\
\mat C^{QT}&\mat C^{QQ}&\mat C^{QU}\\
\mat C^{UT}&\mat C^{UQ}&\mat C^{UU}\\
\end{pmatrix}
\ ,
\end{equation}
which we will assume is composed of a signal and a noise term, $\mat C = \mat S + \mat N$, evaluated for a fiducial model. In Eq. \eqref{ML1b}, $C^r_{\ell}$ is the theoretical CMB power spectrum for mode $r$. If the fiducial $C^r_\ell$ is chosen to be close to the truth, $ C_{ij}$ can be accurately calculated and then Eq.\ \eqref{ML1b}
gives a minimum variance estimate  of the power spectra. One can see from Eq.\ \eqref{eq:cov_QCS}
that the estimator in Eq. \eqref{ML1b} for the $EE$ and $BB$ spectra
mixes $T$ components of the data vector ${\bf x}$  with the much lower amplitude  
$Q$ and $U$ components. This is undesirable, since systematic errors in the $T$ maps and
covariance matrix $\mat C$ could bias the $E$ and $B$ spectra.
We therefore `reshape' $\mat C$ by writing
\begin{equation} \label{eq:cov_recast}
 \tilde{\mat C}=
 \begin{pmatrix}
 \mat C^{TT}&\mat 0&\mat 0\\
 \mat 0&\mat C^{QQ}&\mat C^{QU}\ \\
 \mat 0&\mat C^{UQ}&\mat C^{UU}\\
 \end{pmatrix}
 \ .
\end{equation}
The revised quadratic estimator is then 
\begin{equation}
  \tilde y^r_{\ell} = x_i x_j  \tilde E^{r \ell}_{ij} \ ,  \qquad \tilde{\mat E}^{r\ell}=\frac{1}{2} \tilde{\mat C}\inv \frac{\partial \mat C}{\partial C^{r}_{\ell}}\tilde{\mat C}\inv \ , \label{ML11a}
\end{equation}
with  expectation value
 \begin{equation}
     \langle  y_{\ell}^r\rangle = \tilde{ F}_{\ell \ell^\prime}^{sr}C_{\ell^\prime}^s + \tr\left[\mat N^r\tilde{\mat E}^{r\ell}\right] \ , 
 \end{equation}
Fisher matrix
\begin{equation}
     \tilde{F}^{sr}_{\ell \ell^\prime} = \frac{1}{2} \tr \left[ \frac{\partial \mat C}{\partial C^s_{\ell^\prime}} \tilde{\mat C}\inv \frac{\partial \mat C}{\partial C^r_{\ell}} \tilde{\mat C}\inv \right] \, \label{eq:QCS_fish}
\end{equation}
and variance
\begin{align}
    \langle \tilde y^r_{\ell} \tilde y^s_{\ell^\prime} \rangle - \langle \tilde y^r_{\ell} \rangle \langle y^s_{\ell^\prime}\rangle \equiv  F_{\ell \ell^\prime}^{rs} = 2 \tr \left[ \mat C \tilde{\mat E}^{r \ell} \mat C \tilde{\mat E}^{s \ell^\prime} \right] \ . \label{FF1}
\end{align} 
If the matrix $\mat{\tilde{F}}$ is invertible, unbiased
estimates of the  power spectra can be computed from 
\begin{equation}
    \Hat{C}_{\ell} =  \tilde{ F}\inv_{\ell \ell^\prime} \left(y_{\ell^\prime}-\tr \left[ \mat N\tilde{\mat E}^{\ell^\prime}\right]\right) \ , \label{eq:QCS_true}
\end{equation}
(dropping the mode index $r$ to avoid unwieldy notation in the remainder of this section) with covariance matrix
\begin{equation}
    \langle \Delta \hat{C}_{\ell} \Delta \hat{C}_{\ell^\prime}\rangle = \tilde{\mat F}\inv \mat F \tilde{\mat F}\inv \ .  \label{equ:covest}
\end{equation}
Note that if we had used the exact covariance matrix of Eq. \eqref{eq:cov_QCS}, the Fisher matrix in Eq. \eqref{FF1} would take
the more familiar form 
\begin{equation}
 { F}_{\ell \ell^\prime} = \frac{1}{2} \tr \left[ \frac{\partial \mat C}{\partial C_{\ell}} {\mat C}\inv 
\frac{\partial \mat C}{\partial C_{\ell^\prime}} {\mat C}\inv \right],
\end{equation}
and the covariance matrix of Eq. \eqref{equ:covest} becomes $\langle
\Delta \hat{C}_{\ell} \Delta \hat{C}_{\ell '}\rangle = F\inv_{\, \ell \ell^\prime}$. For
realistic sky cuts, using the reshaped covariance matrix in place of
the exact covariance matrix leads to a negligible increase in variance
compared to the optimal estimator, given in Eq. \eqref{ML1a}
\citep[see][]{2006MNRAS.370..343E}.

In our application, the \Planck noise properties in polarisation are complex
and so it is dangerous to estimate power spectra using Eq. \eqref{eq:QCS_true} since this requires
the subtraction of a noise term. We therefore modified the quadratic estimator
by applying it to cross-spectra of maps $(a)$ and $(b)$ on the assumption that the noise
between these maps is uncorrelated \citep{Efstathiou:2014}. The QCS estimator
is 
\begin{equation}
  \tilde y^{(a,b)}_{\ell} = x^{(a)}_i x^{(b)}_j  \tilde E^{(a, b) \ell}_{ij} \ ,  \ \  \tilde{\mat E}^{(a, b)\ell}=\frac{1}{2} (\tilde{
\mat C}^{(a)})\inv \frac{\partial \mat C}{\partial C_{\ell}}(\tilde{\mat C}^{(b)})\inv \ , \label{equ:QCS0}
\end{equation}
which can be computed rapidly via spherical harmonic transforms \citep{2006MNRAS.370..343E}.
For each map, ${\bf x}^{(k)}$, we form the weighted map 
\begin{equation}
{\bf z}^{(k)}_i = (\tilde{\mat C}^{(k)})\inv {\bf x}^{(k)}, \label{equ:QCS1}
\end{equation}
and compute the PCL cross spectrum $C^{z(a,b)}_\ell$ of the maps ${\bf z}^{(a)}$ and ${\bf z}^{(b)}$. The estimator
in Eq. \eqref{equ:QCS0} is then
\begin{equation}
   {\tilde y}^{(a,b)}_{\ell} = \frac{(2\ell +1)}{2 \Omega^2} \tilde{C}^{z(a,b)}_{\ell} \ ,  \label{equ:QCS2}
\end{equation}
where $\Omega$ is the solid angle of a single map pixel, each assumed to be of the same size. The QCS estimates
can therefore be computed very rapidly for large numbers of simulations, since the matrices $(\tilde{\mat C}^{(a)})\inv$
and $(\tilde{\mat C}^{(b)})\inv$ need only be computed once.

The expectation value
of Eq. \eqref{equ:QCS0} is 
 \begin{align}
     \langle  y_{\ell}^{(a, b)}\rangle &= \tilde{F}^{(a,b)}_{\ell \ell^\prime} C_{\ell^\prime}  ,  \\    \tilde{ F}^{(a,b)}_{\ell \ell^\prime} &= \frac{1}{2} \tr \left[ \frac{\partial \mat C}{\partial C_{\ell}} (\tilde{\mat C}^{(a)})\inv 
\frac{\partial \mat C}{\partial C_{\ell^\prime}} (\tilde{\mat C}^{(b)}) \inv \right], \label{equ:QCS3}
 \end{align}
and estimates of the power spectra $\Hat C_\ell$ can be recovered by inversion of Eq. \eqref{equ:QCS3} as in Eq. \eqref{eq:QCS_true}.

The QCS estimator was used, together with the simulation-based likelihood (\simbal), 
 to analyse the \Planck HFI maps in \citetalias{2016A&A...596A.107P}, \citetalias{Aghanim:2018eyx} and 
\citet{2019arXiv190809856P}. Although the QCS estimator is not `optimal' in any formal sense, it has a  significantly lower variance than
a PCL estimator applied to the \Planck polarisation maps. However in addition to lower variance, the QCS estimator produces
estimates of the $EE$ power spectrum with a covariance matrix that is effectively diagonal. It is because the QCS estimates
of $\hat C^{EE}_\ell$ for each multipole are effectively independent that the \simbal likelihood approach is feasible. 

The variance of the QCS estimates is somewhat more complicated than Eq. \eqref{FF1}:
\begin{align} \label{equ:QCS4} 
   \langle  &y^{(a,b)}_{\ell}  y^{(a,b)}_{\ell '} \rangle - \langle y^{(a,b)}_{\ell} \rangle \langle y^{(a,b)}_{\ell '} \rangle =  \\ 
    &\left[ 2 S_{ip} S_{jq} + \left(  N_{ip}^{(a)} +  N_{ip}^{(b)} \right) S_{jq} +  N_{ip}^{(a)} N_{jq}^{(b)} \right]   \tilde { E}_{ij}^{(a, b)\ell}  \tilde { E}_{pq}^{(a, b)\ell^\prime}\ . \nonumber
\end{align}
The error bars shown in plots of the power spectra below (Figs. \ref{fig:QCS_100x143_data} and \ref{fig:summary_QCS_cross_check}) are computed from Eq. \eqref{equ:QCS4}, though this expression is not used in the likelihoods. In this paper, we assume a base $\Lambda$CDM model with $\tau = 0.06$ to compute the signal matrix ${\bf S}$. The construction of the noise covariance matrices ${\bf N}$ (including realistic correlated noise)  is described in Sec. \ref{sec:NCM} and is based on the end-to-end simulations described in \cite{HFI_data:2018} and \citetalias{2019A&A...629A..38D}, respectively, for the analysis of \SRone and \SRtwo maps. In contrast, the analysis of \cite{2019arXiv190809856P} used a simplified noise model based on the \Planck FFP8 simulations \citep{2016A&A...594A..12P} for the QCS computations.

\section{Data} \label{sec:data}

The \SRone and \SRtwo map-making algorithms are described in detail in \citetalias{2016A&A...596A.107P}, \cite{HFI_data:2018} and \citetalias{2019A&A...629A..38D}.
Briefly, the algorithms find global solutions minimising the variance in the response of each polarised bolometer 
within a given frequency band with respect to a number of instrumental parameters. Analogue to digital
converter nonlinearity (ADCNL) introduced large polarisation systematics at low multipoles 
in the 2015 \Planck HFI maps \citep{2016A&A...594A...8P}. These systematics were substantially
reduced with the \SRone processing used to produce the \Planck 2018 HFI legacy maps. Although \SRone reduced systematics
arising from first order ADCNL,  second order ADCNL caused temperature to polarisation dipole leakage in the maps \citep{Aghanim:2018eyx}. The \SRtwo map-making algorithm reduced these large scale polarisation systematics for $100$ and $143$ GHz still further via the following refinements: 
\begin{enumerate}
    \item the revised ADCNL corrections in \SRtwo obviate the need for fitting an effective gain variation of the bolometers;
    \item the polarisation angle and efficiency for each bolometer were treated as marginalised parameters;
    \item the thermal dust and CO templates were updated. 
\end{enumerate}
These improvements reduced significantly large-scale systematics in the polarisation data. The \SRone and \SRtwo $100$ and $143$ GHz Q and U maps are compared in Figs. 4 and 5 of \cite{2019arXiv190809856P}.
In this paper, we have analysed both the \SRone and \SRtwo maps, together with their respective sets of
 end-to-end simulations, so that the
reader can assess the impact of changes in the HFI data processing.

\subsection{Map compression and foreground cleaning}\label{sec:CMB_data}
\begin{figure}
    \centering
     \includegraphics[width=\columnwidth]{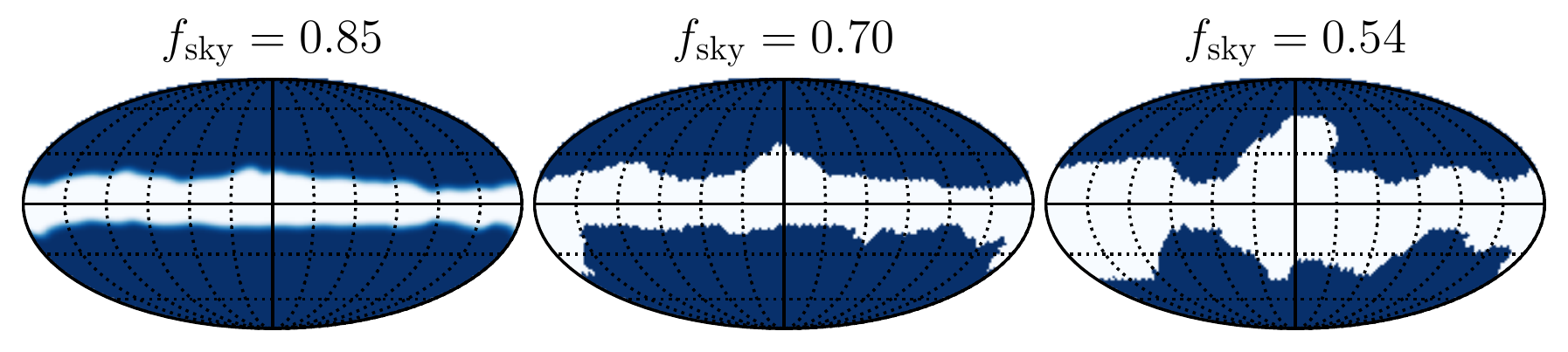}
    \vspace{-0.2in}
    \caption[Masks.]{Masks used in this paper. The quantity $f_{\rm sky}$ denotes the fraction of masked pixels, $f_{\rm sky} =\sum_{i=1}^{N_{\rm pix}} x_i /N_{\rm pix}^{\rm tot}$. The apodised mask ($0\leq x_i \leq 1$) with $f_{\rm sky}=0.85$ is used as a `processing' mask to degrade the high resolution maps to low resolution (as discussed in the text). The mask with $f_{\rm sky}=0.70$ is binary and used to compute foreground cleaning coefficients and to mask the temperature maps when computing power spectra. The more conservative mask with $f_{\rm sky}=0.54$  is binary as well and used to compute polarisation power spectra.}
    \label{fig:masks}
    \vspace{-0.1in}
\end{figure}

To apply the QCS estimator,  we  degrade the high resolution \Planck $T$, $Q$ and $U$ maps following a
 procedure similar to that described in \citetalias{2016A&A...596A.107P} and \citetalias{2019A&A...629A..38D}. 
We first apply an apodised mask with $f_{\rm sky}=0.85$ (as plotted in Fig. \ref{fig:masks}) to suppress the Galactic plane region\footnote{This procedure is unnecessary for foreground subtracted simulations, but required for the real data 
 to avoid smearing high amplitude foreground emission in the Galactic plane to high Galactic latitudes.}. 
The maps were then smoothed  using the harmonic-space smoothing operator:
\begin{equation} \label{eq:harmonic_space_operator}
w(\ell) = \left\{ \begin{array}{ll}
 1 &,\ \ell \leq \ell_1\\
 \frac{1}{2}\left[ 1 + \cos{\pi\frac{\ell - \ell_1}{\ell_2-\ell_1}} \right] &,\ \ell_1 < \ell \leq \ell_2 \ ,\\
 0 &,\ \ell > \ell_2
\end{array} \right. \end{equation}
with $\ell_1 = N^{\rm lr}_{\rm side}$ and $\ell_2 = 3N^{\rm lr}_{\rm side}$, and degraded from $N_{\rm side}= 2048$ ($5.03 \times 10^7$ 
\texttt{HealPix} pixels, \cite{2005ApJ...622..759G})  to  $N^{\rm lr}_{\rm side}= 16$ ($3072$ pixels) in the low resolution maps. We apply the smoothing operator given in Eq. \eqref{eq:harmonic_space_operator} and a $\texttt{HealPix}$ pixel window function at the map level to match the $N^{\rm lr}_{\rm side}= 16$ low resolution covariance matrices, $\mat N_0$,  discussed in Sec. \ref{sec:NCM}.

The low resolution maps are foreground cleaned by fitting  high and low frequency templates. Specifically, 
we use the $353$ GHz maps as a dust template and (in polarisation only) either the \Planck LFI
$30$ GHz or \WMAP K-band (22 GHz) maps as synchrotron templates.  Following \citet[][hereafter \citetalias{EG19}, see Secs. 7.1 and 8]{EG19} we minimise cleaned map residuals
\begin{equation}
 \sigma^2 = \sum_i ( (1 + \alpha_1 + \alpha_2) m_i -  \alpha_1 m^{T_1}_i - \alpha_2 m^{T_2}_i)^2 , \label{eq:clean1} 
\end{equation}
with respect to the coefficients $\alpha_1$ and $\alpha_2$ for the two map templates
$m^{T_1}_i$ and  $m^{T_2}_i$. In polarisation, the sum in Eq. \eqref{eq:clean1} extends over the unmasked
pixels in the $Q$ and $U$ maps defined by the $f_{\rm sky} = 0.70$ mask plotted in Fig. \ref{fig:masks}. We therefore
 determine two sets of coefficients for polarisation, which we denote
$\alpha^P_1$ and $\alpha^P_2$. For temperature, we have applied dust template subtraction with a coefficient
$\alpha^T_1$ and ignored synchrotron,  for reasons discussed below.

\begin{table}
\centering
{\def\arraystretch{1.3}\tabcolsep=5pt
\begin{tabular}{lllllll}
\hline \hline
  Data set & $\nu$ [GHz]& $\alpha_{353}^T$ &   $\alpha_{353}^P$ & $\alpha_{30}^P$&  $\alpha_{353}^P$ & $\alpha^P_{22}$  \\ \hline
\SRone  & 100 & 0.0237 & 0.0190 & 0.0126 & 0.0179 & 0.0104 \\ 
        & 143 & 0.0398 & 0.0402 & 0.0096 & 0.0396 & 0.0071 \\\hline
\SRtwo  & 100 & 0.0237 & 0.0193 & 0.0191 & 0.0189 & 0.0094 \\ 
        & 143 & 0.0398 & 0.0396 & 0.0102 & 0.0391 & 0.0062 \\ \hline
\end{tabular}}
\caption{Cleaning  coefficients at low multipoles for temperature and polarisation maps. The polarisation cleaning coefficients are listed in pairs, depending on whether \Planck 30 GHz ($\alpha^P_{30}$) or \WMAP 22 GHz ($\alpha^P_{22}$) polarisation maps were used 
as synchrotron templates. In temperature, we cleaned only for dust using the 353 cleaning coefficients listed in bold face from Table 7 of \citetalias{EG19}.}
\label{tab:cleaning_coeff_353}
\vspace{-0.1in}
\end{table}

The template coefficients used in this paper are listed in Table \ref{tab:cleaning_coeff_353}.
The polarisation coefficients are listed in pairs, one pair for each of the \SRone and \SRtwo maps to give
an impression of the sensitivity of the dust coefficient on the choice of low frequency template. Polarised
dust emission dominates the $100$ and $143$ GHz  $Q$ and $U$ maps at low resolution,  with synchrotron making
a small (but non-negligible) contribution at $100$ GHz. As
Table \ref{tab:cleaning_coeff_353} shows, the polarisation dust coefficients are stable. However, for $100$ GHz
the amplitude of the $30$ GHz coefficient differs  between \SRone and \SRtwo.  The \SRtwo polarisation 
cleaning coefficients are in excellent agreement with the coefficients determined by \cite{2019arXiv190809856P}
($\alpha^P_{353} = 0.0186$, $\alpha^P_{22} = 0.0095$ for $100$ GHz). As we will show below, ignoring the
synchrotron correction in polarisation causes shifts in $\tau$ of a fraction of a standard deviation. Maps cleaned with \WMAP K-band are nearly indistinguishable to the ones cleaned with \Planck 30 GHz maps. This similarity gives us confidence in our synchrotron cleaning coefficients.

Dust cleaning in temperature using $353$ GHz or higher frequencies removes almost all of the foreground emission at low multipoles  at $143$ GHz leaving noise-free CMB signal over most of the sky (and  indistinguishable from the \Planck component separated maps), as discussed in detail in \citetalias{EG19}.  At $100$ GHz, the main contaminant,  following $353$ cleaning,  is CO line emission which 
makes a small but easily detectable contribution to the signal. Since the residual foregrounds are small at low multipoles, we
subtract only dust emission in temperature using $353$ GHz maps and the cleaning coefficients determined by \citetalias{EG19} (as  
listed in Table \ref{tab:cleaning_coeff_353}). We use the same temperature cleaning coefficients for \SRone and \SRtwo since these coefficients are insensitive to the map-making algorithm.

To avoid introducing correlated noise into the QCS spectra, we foreground clean the frequency maps using pairs of half-mission (HM) template maps\footnote{For HFI, the half-mission maps are constructed by splitting the available rings from the full-mission frequency maps into two halves, \ie each frequency channel has two half-mission maps: HM1 and HM2}. For example, for the $100\times143$ cross-spectrum, we clean the 100 GHz channel with 353 GHz HM1 and 30 GHz HM1 maps and the 143 GHz channel with 353 GHz HM2 and 30 GHz HM2 maps.

\begin{figure}
    \centering
    \includegraphics[width=\columnwidth]{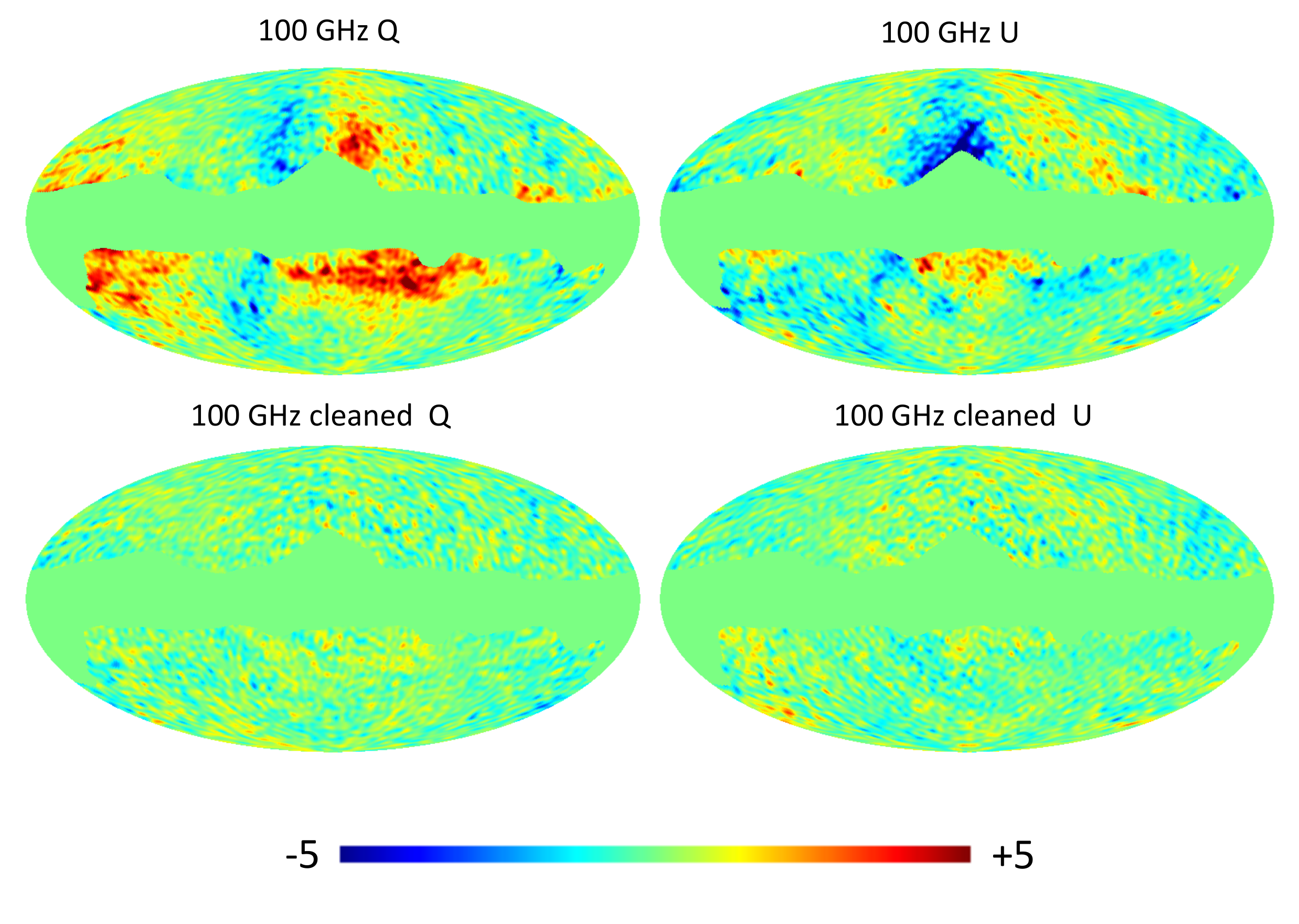} \\
    \includegraphics[width=\columnwidth]{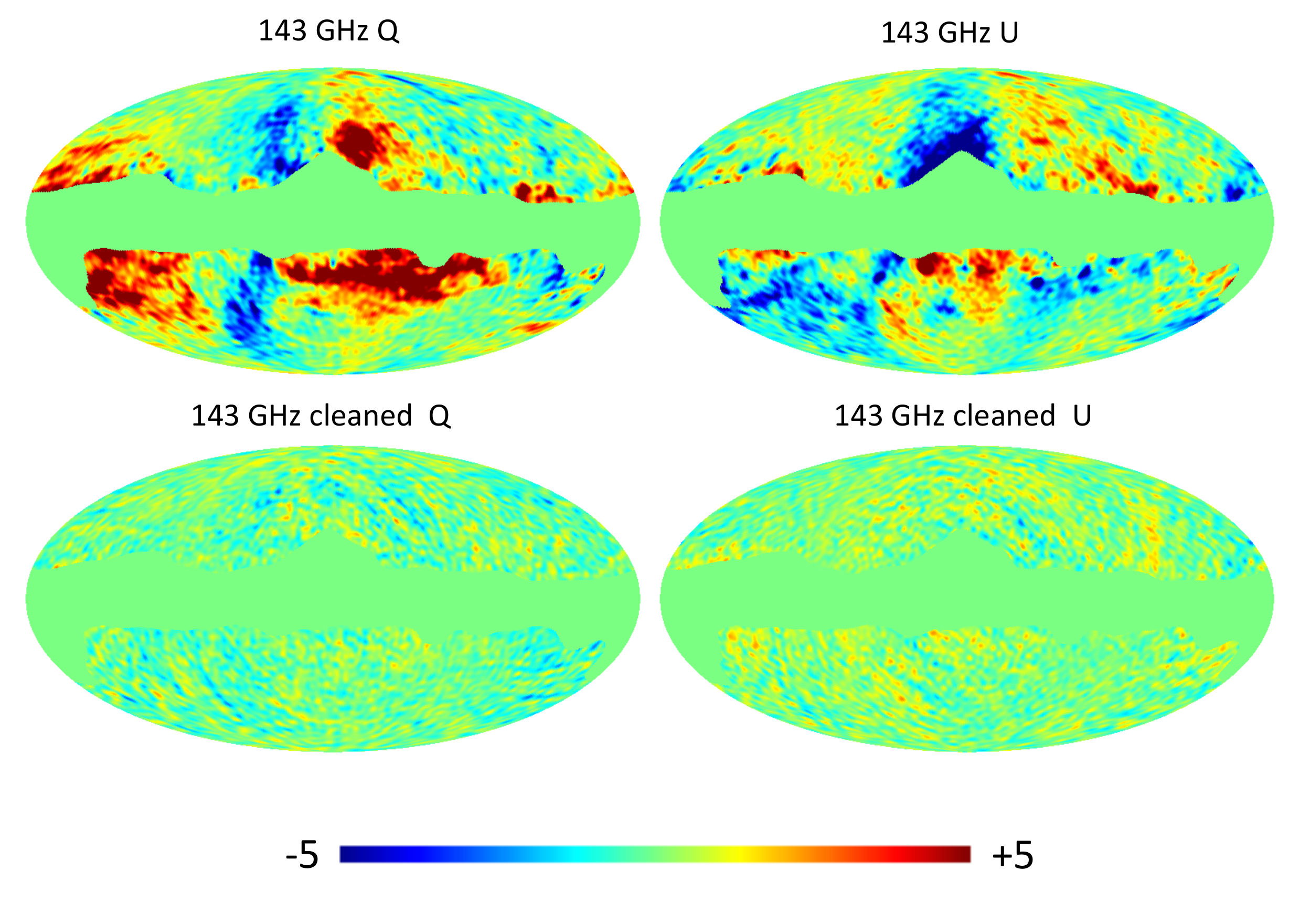} \\
    
    \vspace{0.0in}
    \caption{Full-mission \SRtwo $Q$ and $U$ maps at 100 and 143 GHz 
degraded to $N_{\rm side}=128$ smoothed with a $\sigma = 2^{\circ}$ Gaussian beam. The  $f_{\rm sky}=0.70$ mask of Fig. \ref{fig:masks}
has been applied. The upper set of maps at each frequency shows the $Q$ and $U$ maps before foreground subtraction.
The lower set of maps at each frequency shows the $Q$ and $U$ maps after subtraction of $353$ and $30$ GHz \Planck maps
with the template cleaning coefficients listed in Table \ref{tab:cleaning_coeff_353}. The colour scale is in units of $\mu{\rm K}$.}
    \label{fig:cleaned_maps}
    \vspace{-0.1in}
\end{figure}

\begin{figure}
    \centering
    \includegraphics[width=\columnwidth]{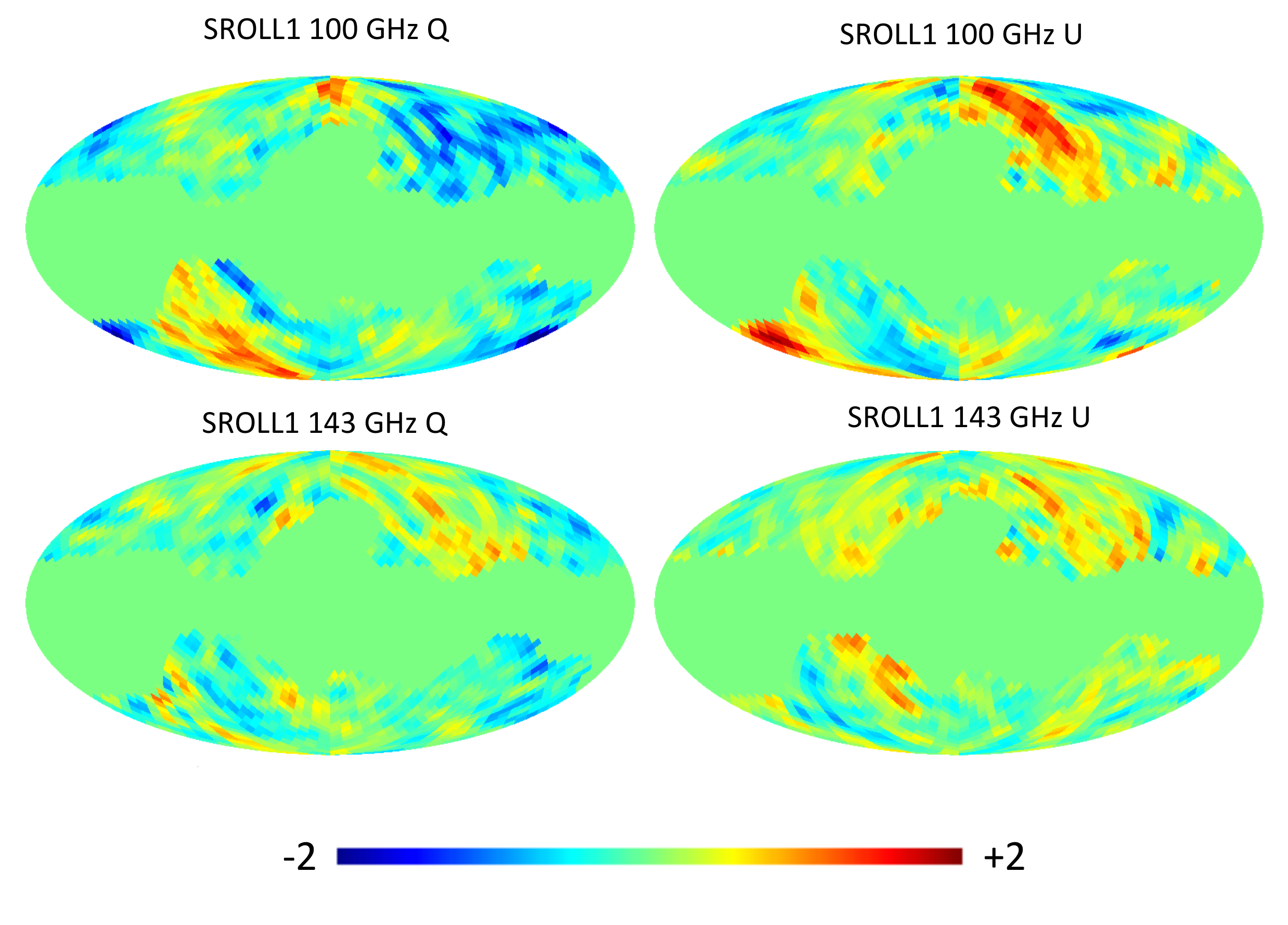} \\
    \includegraphics[width=\columnwidth]{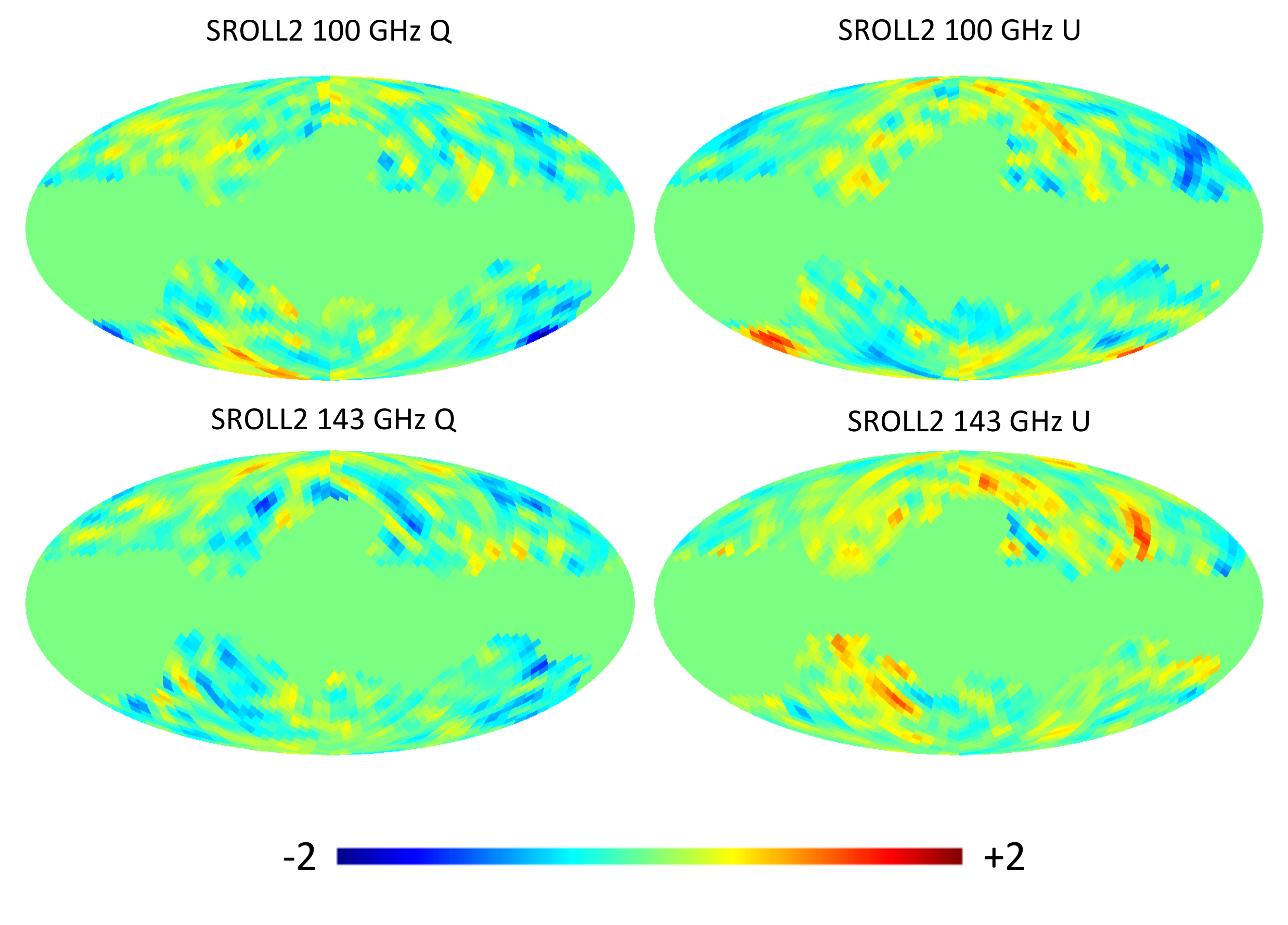} \\

    \vspace{-0.049in}
    \caption{Polarisation maps used for the main cosmological results presented in this paper.
The plots show the $N^{\rm lr}_{\rm side} = 16$ foreground cleaned full-mission  $100$ and $143$ GHz $Q$ and $U$ 
maps. \SRone and \SRtwo maps are shown in the upper and lower panels respectively. The $f_{\rm sky} = 0.54$ mask of
Fig. \ref{fig:masks} has been applied. The colour scale is in units of $\mu{\rm K}$.}
    \label{fig:cleaned_maps2}
    \vspace{-0.1in}
\end{figure}

The \SRtwo polarisation maps before and after foreground cleaning are plotted in Fig. \ref{fig:cleaned_maps}, which shows clearly that the 
polarisation maps at $100$ and $143$ GHz are dust dominated at low multipoles before foreground subtraction. Fig. \ref{fig:cleaned_maps2}
shows the foreground cleaned $N^{\rm lr}_{\rm side} = 16$ $Q$ and $U$ maps used for the main cosmological results in this paper. The \SRone
maps are shown in the upper panels and the \SRtwo maps are shown in the lower panels. One can clearly see systematic features in the 
\SRone maps, particularly at $100$ GHz. These systematic features can be reduced  by subtracting a smoothed mean map $\mathbf{\Bar{n}}$
determined from end-to-end simulations as discussed in the next section. For reference, the rms fluctuations in the \SRtwo $100$ GHz polarisation maps
over the unmasked area shown in Fig. \ref{fig:cleaned_maps2} are $\sigma^P = 0.66 \ \mu{\rm K}$ before foreground cleaning and
 $\sigma^P = 0.36 \ \mu{\rm K}$ after foreground cleaning. The rms of the scaled 353 GHz dust template is $\sigma^P = 0.55 \  \mu{\rm K}$ and for the 
scaled 30 GHz synchrotron template is $\sigma^P = 0.11 \ \mu{\rm K}$.

\subsection{Noise covariance matrices}\label{sec:NCM}
In this section, we describe how we fit a parametric model to estimates of the pixel-pixel
noise covariance matrices (NCMs)  computed from end-to-end simulations. Once we have a model for a NCM, 
we can generate large numbers of simulations on the assumption that the noise is Gaussian, allowing us to 
construct likelihoods as discussed in Sec.\ \ref{sec:Likelihood_sec}. 
We compute two sets of NCMs, one each 
for \SRone and \SRtwo, from the end-to-end simulations of the 
 map-making pipelines\footnote{\SRone simulations (labelled FFP10) available at \url{https://pla.esac.esa.int};
  \SRtwo simulations available at \url{http://sroll20.ias.u-psud.fr}.}.  In each case, the fiducial CMB and the input foreground model was subtracted,  leaving maps containing noise and map-making systematics.   We then construct 
empirical  NCMs, $\mathbf{\Hat{N}}$, 
\begin{equation} \label{eq:NCM}
    \mathbf{\Hat{N}} = \frac{1}{n_s} \sum_{i=1}^{n_s} (\mathbf{n}_i - \mathbf{\Bar{n}})(\mathbf{n}_i - \mathbf{\Bar{n}})^{\top} \ ,
\end{equation}
where $\mathbf{n}_i$ are the simulated sky maps for the Stokes parameters $T,Q$ and $U$, $\mathbf{\Bar{n}}$ a smoothed template of the mean of the  maps (see Eq. \eqref{eq:smooth_template}) and $n_s$ the number of 
simulations used in the sum in Eq. \eqref{eq:NCM}.  To test whether our methodology is biased as a consequence of 
overfitting  we do not use all of the end-to-end simulations to compute Eq. \eqref{eq:NCM}. We excluded 
 $100$ simulations from each set to allow validation tests of the likelihoods  as discussed in  
 Sec.\ \ref{sec:compare_likelihoods}. 

We approach the problem of fitting a model to $\mathbf{\Hat{N}}$  as a maximum likelihood inference problem. We assume a Gaussian probability distribution for each noise realisation
\begin{align}
    \mathcal{L} \equiv \mathbf{\mathcal{P}}(\mathbf{\hat{N}} \vert \mathcal{\mathbf{M}}) &= \prod_{i=1}^{n_s} \frac{1}{\sqrt{|2\pi \mathbf{M}|}}e^{-\frac{1}{2}(\mathbf{n}_i - \mathbf{\Bar{n}}) \mathbf{M}\inv(\mathbf{n}_i-\mathbf{\Bar{n}})^{\top}} \nonumber\\ &= \frac{1}{|2\pi \mathbf{M}|^{\frac{n_s}{2}}} e^{-\frac{n_s}{2}\tr \ \mathbf{M}\inv \mathbf{\hat{N}}}  \ ,
\end{align}
where  $\mat M$  is the model for the  noise covariance matrix. We assume that $\mat M$ consists of three terms
\begin{equation} \label{eq:noise_fit}
    \mat M  =  \alpha \mat N_0 + \beta \mat N_1 + \mat{Y\mathbf{\Psi}Y}^\top, 
\end{equation}
where $\alpha$ and $\beta$ are scaling parameters for two matrices $\mat N_{0}$ and $\mat N_1$ and the term
$\mat{Y \mathbf{\Psi}Y}^\top$ models large-scale modes with parameters $\mathbf{\Psi}$ as described more fully below. Since we are dealing with
$Q$ and $U$ maps, the noise covariance matrices $\mat M$ are of size $(2N_{\rm pix},2N_{\rm pix})$. We neglect noise 
in temperature and only fit polarisation noise, since the low resolution $T$ maps are signal dominated to high accuracy. Moreover, we assume that the noise for $100$ and $143$ GHz maps is uncorrelated.

The matrices $\mat N_0$ are the $N^{\rm lr}_{\rm side} = 16$ low resolution
map-making covariance matrices \citep [see e.g][]{Tristram:2011} computed for the
parameters of the FFP8 simulations \citep{2016A&A...594A...8P,
  2016A&A...594A..12P}. These covariance matrices contain structure
representing the scanning strategy, detector white noise and `$1/f$'-type 
noise, but do not include complexities associated with corrections
for ADCNL. Since these matrices were designed to match the 2015
\Planck half-mission maps, they do not necessarily match the noise
levels of the \SRone and \SRtwo maps at high multipoles. The matrices
$\mat N_1$ were constructed from the diagonal components ($(\sigma^2)^T_i$, 
$(\sigma^2)^Q_i$, $(\sigma^2)^U_i$  of  the
$3\times3$ $T, Q, U$ high resolution pixel noise estimates produced by
the map-making algorithms. These noise estimates were degraded in resolution
to $N^{\rm lr}_{\rm side} = 16$ (appropriate for a low resolution map $X_i$) by computing:
\begin{align} \label{eq:low_res_cov}
    \mat N_{\rm 1} \equiv \langle \mat X_i \mat X_j \rangle = & \sum_{\ell_1\ell_2} \sum_{pq}  \sigma^2_p \delta_{pq} \frac{(2\ell_1+1)}{4\pi} \frac{(2\ell_2+1)}{4\pi} \Omega_p\Omega_q \nonumber 
\\ & \qquad \times P_{\ell_1}(\cos{\theta_{ip}})P_{\ell_2}(\cos{\theta_{jq}})f_{\ell_1}f_{\ell_2} \ ,
\end{align}
for each  $Q$ and $U$  (see App. A of \cite{Efstathiou:2009}). We assume that the noise is diagonal $\langle x_ix_j \rangle = \sigma^2_p \delta_{pq}$, $\Omega_i$ is the solid angle of a high-resolution map pixel, $P_{\ell}$ denotes the Legendre polynomials and $f_{\ell_i}$ is the smoothing operator applied to the high-resolution maps given in Eq. \eqref{eq:harmonic_space_operator}. The sum
\begin{equation}
\mat N = \alpha \mat N_0 + \beta \mat N_1
\end{equation}
therefore allows us to model destriping noise correlations with adjustable `white noise' levels, using the matrices $\mat N_0$
and $\mat N_1$ as templates.

The final term in Eq. \eqref{eq:noise_fit} models additional large-scale noise correlations in the \SRone and
$\SRtwo$ end-to-end simulations.  Terms up to $\ell_{\rm max} = 4$ are added to the noise model via 
$\mat Y \mathbf{\Psi} \mat Y^{\top}$, where each column of the matrix $\mathbf{Y}$ is a map of the spherical harmonic functions $Y_{\ell m}(\theta, \varphi)$. The square matrix $\mathbf{\Psi}$, with dimensions equal to the number of large-scale modes that we wish to fit ($42$ modes to model $QQ$, $QU$ and $UU$ correlations), controls the covariance of the modes. In the noise fitting procedure we apply the binary polarisation mask with $f_{\rm sky}=0.54$, shown in Fig. \ref{fig:masks}, to the $Q$ and $U$ maps. This avoids fitting modes behind the Galactic plane. 

To solve our inference problem, we minimise the `action' $\mathcal{S} \equiv -\ln(\mathbf{\mathcal{P}}(\mathbf{\hat{N}} \vert \mathcal{\mathbf{M}}))$ 
\begin{equation}
    \mathcal{S} = \frac{n_s}{2}\left(\tr \ \mathbf{M}\inv \mathbf{\hat{N}} + \ln{|\mathbf{M}|}\right) \ \label{eq:sfornoise}
\end{equation}
 with respect to the free parameters in our model for $\mat M$.
We solve numerically for the scalar parameters $\alpha$ and $\beta$, solving analytically for the $\mathbf{\Psi}$ at each step. 
The matrix $\mathbf{\Psi}$ is given by (for the derivation see App. \ref{sec:analytic_solution})
\begin{equation}
   \mathbf{\Psi} = (\mat Y^\top \mat N\inv \mat Y)\inv \left( \mat Y^\top \mat N\inv \left[ \mat{\Hat{N}}- \mat M \right] \mat N\inv \mat Y\right) (\mat Y^\top \mat N\inv \mat Y)\inv \ . \label{eq:solnforpsi}
\end{equation}
When fitting the NCM $\mat M$ using Eq. \eqref{eq:noise_fit} the dominating components are the $1/f$-type noise matrix ($\alpha\approx 1.8$) as well as the large scale modes up to $\ell_{\rm max}=4$. The smoothed low resolution covariance matrix
$\mat N_{\rm lr}$ only subtracted a small amount of power from the diagonal with $\beta\approx-0.1$.

As summarised in the start of this section,  uncorrected ADCNL 
leads to `stripy' residuals in the \Planck $Q$ and $U$ maps  at 100, 143, and 217 GHz, which are
substantially reduced in \SRtwo compared to \SRone \citep[see, for example, Fig. 4 in][]{2019arXiv190809856P}. The end-to-end
simulations provide templates for these residuals. Instead of subtracting the mean map averaged over the simulations, $\Bar{\mat x}$, we subtract a smoothed template 
\begin{equation} \label{eq:smooth_template}
    \Bar{\mat n}  = \mat Y (\mat Y^\top \mat N_0\inv \mat Y)\inv \mat Y^\top \mat N_0\inv \Bar{\mat x} \ ,
\end{equation}
with $\ell_{\rm max} = 4$ (this is the maximum likelihood solution of the map-making equation, as discussed 
in \eg \cite{Tegmark:1997}). We have approximated the NCM in Eq.  \eqref{eq:smooth_template} by the term $\mathbf{N_0}$ to avoid
having to iterate to obtain a solution for $\mathbf{M}$.

\subsection{Quadratic temperature and polarisation power spectra}
\begin{figure} 
    \centering
    \includegraphics[width=\columnwidth]{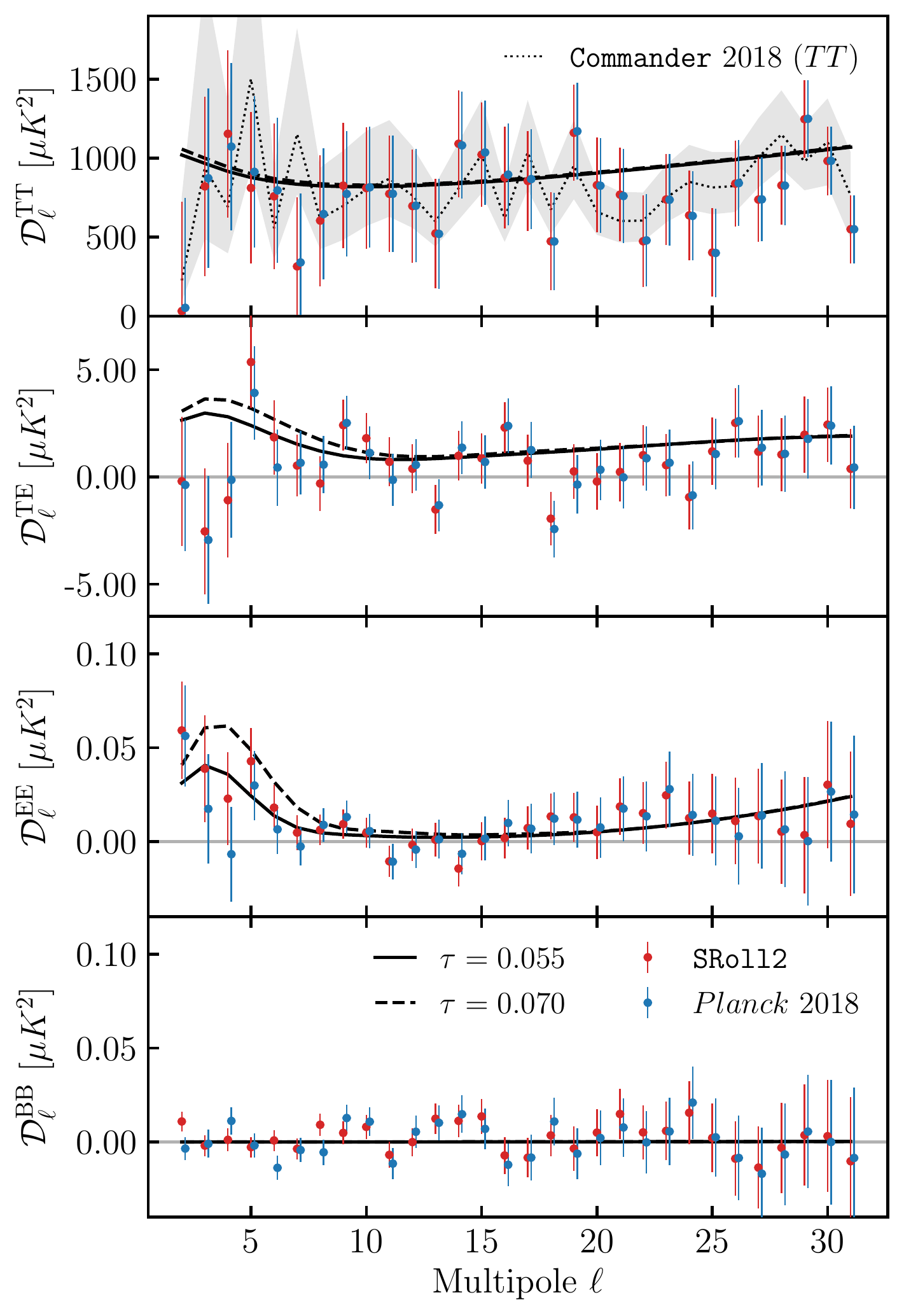}
    \vspace{-0.2in}
    \caption[QCS estimates of \SRtwo maps.]{Low-$\ell$ QCS estimates for $100\times143$ temperature ($TT$, $TE$) and polarisation ($EE$, $BB$) full-mission cross-spectra of \SRtwo (red) and \Planck 2018 (blue) maps with $\mathcal{D}_{\ell} = \ell (\ell +1) C_{\ell}/(2 \pi)$. The \commander 2018 $TT$ cross-spectrum \citepalias{Aghanim:2018eyx} (dotted lined) with the associated error bar as a grey shaded region is shown for comparison. The maps are foreground cleaned with the 30 GHz and 353 GHz channel for synchrotron and dust emission, respectively. Theoretical spectra are shown for $\tau = 0.055$ (solid line) and $\tau=0.070$ (dashed line).} 
    \label{fig:QCS_100x143_data}
    \vspace{-0.1in}
\end{figure}

As an illustration of our methods, Fig.\ \ref{fig:QCS_100x143_data} shows the $TT$, $TE$, $EE$ and $BB$ spectra, computed using the QCS estimator technique presented in Sec.\ \ref{sec:CMB_QCS_estimator}, for the cross correlation of the $100$ and $143$ GHz full-mission maps for both \SRone and \SRtwo for $2\leq \ell \leq 31$. The $T$, $Q$ and $U$ maps have been foreground cleaned as discussed in Sec.\ \ref{sec:CMB_data}. We apply a mask with $f_{\rm sky}=0.70$, shown in Fig.\ \ref{fig:masks}, to the $T$ maps and we apply a mask with $f_{\rm sky}=0.54$, also shown in Fig.\ \ref{fig:masks}, to the $Q$ and $U$ maps. For the $TT$ cross-spectra we compare our results with the \commander 2018 spectrum \citepalias{Aghanim:2018eyx} and observe good agreement between them, even though the \commander $TT$ spectrum is computed using a mask with larger sky fraction ($f_{\rm sky}=0.86$). In particular, we observe the same behaviour in the spectra: first, at $\ell=2$ there is a very low value, and, second, there is a  `dip' at around $\ell \simeq 20-25$. The error bars on the \commander spectrum are asymmetric $1\sigma$ posterior widths. 

To guide the eye, the solid and dashed black lines in Fig. \ref{fig:QCS_100x143_data} show the $TT$, $TE$ and $EE$ spectra for base $\Lambda$CDM model with $\tau=0.055$, close to the best fit value for the \SRone and \SRtwo maps cited in Eqs. \eqref{eq:Tau1} and \eqref{eq:Tau2}, and for $\tau = 0.070$, which is disfavoured at about the  $2-3 \sigma$ level. The $TT$ and $TE$ spectra have large cosmic variance and so do not provide strong constraints on $\tau$ given the low values of $\tau$ inferred from the $EE$ spectra. 
The \SRone and \SRtwo $TT$, $TE$ and $EE$ spectra are very similar and mainly show differences at low multipoles. The $BB$ spectra are approximately consistent with zero, providing an important null-test for \SRone and \SRtwo. The $\chi^2$ values divided by the degrees of freedom (29 multipoles) are 1.28 for \SRone and 1.13 for \SRtwo.

\begin{figure} 
    \centering
        \includegraphics[width=\columnwidth]{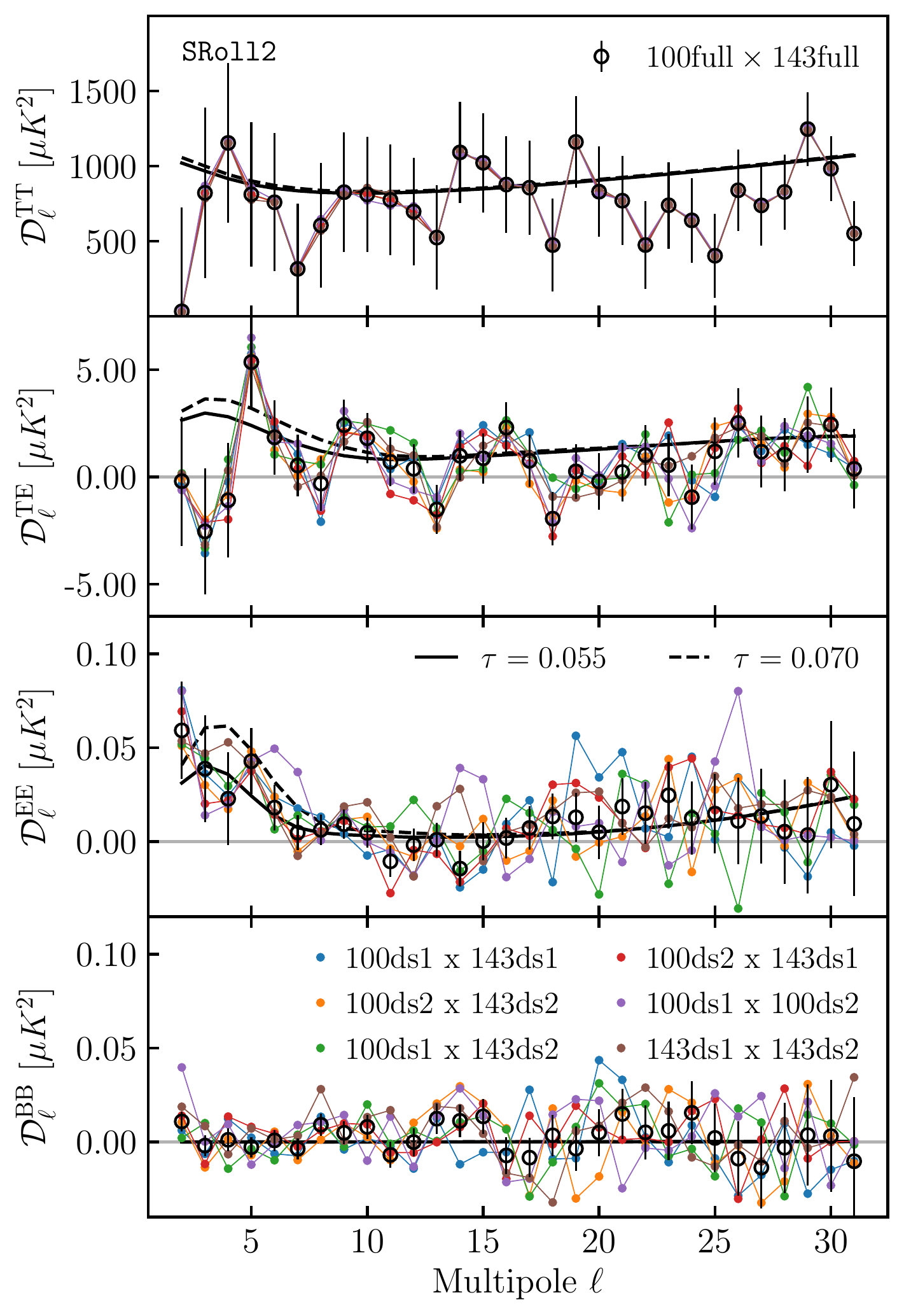}
    \vspace{-0.2in}
    \caption[QCS estimates of \SRtwo maps.]{Low-$\ell$ QCS estimates for $TT$, $TE$, $EE$ and $BB$ auto- and cross-spectra of detector set combinations for 100 and 143 GHz low-resolution \SRtwo maps with $\mathcal{D}_{\ell} = \ell (\ell +1) C_{\ell}/(2 \pi)$. The maps are foreground cleaned with the 30 GHz and 353 GHz channels for synchrotron and dust emission, respectively. Theoretical spectra are shown for $\tau = 0.055$ (solid line) and $\tau=0.070$ (dashed line). For comparison, the black circles show the $100\times143$ full-mission QCS estimates with corresponding error bars.}
    \label{fig:summary_QCS_cross_check}
    \vspace{-0.1in}
\end{figure}

Fig. \ref{fig:summary_QCS_cross_check} shows $TT$, $TE$, $EE$ and $BB$ spectra computed using the QCS scheme for intra- and inter-frequency cross-spectra for combinations of the publicly available \SRtwo detector set\footnote{A detector set denotes a combination of bolometers chosen able to fully determine the polarisation state of (non-circularly-polarised) incoming light: At 100 GHz, ds1 consists of the 100-1a/b and 100-4a/b bolometer pairs and ds2 of the 100-2a/b and 100-3a/b pairs. At 143 GHz, ds1 consists of the 143-1a/b and 143-3a/b pairs, and ds2 of the 143-2a/b and 143-4a/b pairs \citepalias{2016A&A...596A.107P}.} maps. The $T$, $Q$ and $U$ maps are foreground cleaned, following the procedure discussed in Sec. \ref{sec:data}, \ie the ds1 maps are cleaned using 353 GHz HM1 and 30 GHz HM1 maps and the ds2 maps are cleaned with 353 GHz HM2 and 30 GHz HM2 maps. The temperature and polarisation masks are the same as those used as for the previously discussed full-mission cross-spectra and shown in Fig.\ \ref{fig:masks}. For comparison, the empty black circles are the $100\times143$ full-mission QCS estimates with the corresponding error bars (as in Fig. \ref{fig:QCS_100x143_data}).

For the signal-dominated $TT$ and $TE$ cases seen in Fig. \ref{fig:summary_QCS_cross_check}, we measure almost identical spectra for the different detector set combinations as expected. In the low multipole $EE$ polarisation regime  that is most constraining for $\tau$ ($2\leq \ell \lesssim 12$), the estimates scatter between the theoretical curves with $\tau=0.055$ and $\tau=0.070$. This indicates that higher values of $\tau$ may be favoured for some detector set combinations. At intermediate scales, $10 \leq \ell \leq 25$, the spectra scatter around the theoretical curves and no clear trend is visible. However, the $100\ \mathrm{ds1}\times100\ \mathrm{ds2}$ intra-frequency cross spectrum shows a few outliers in polarisation (purple line). This potentially suggests that the 100 GHz maps are more affected by noise and unresolved systematics than the 143 GHz ones. The illustrated $BB$ spectra are all approximately consistent with the null hypothesis of zero signal power.

\section{Likelihoods} \label{sec:Likelihood_sec}
In this section we present and compare a simulation-based likelihood, a likelihood-approximation scheme and a likelihood-free approach. The methods are called: \simbal, \glass and \delfi and their respective likelihoods are named \simlow\footnote{We implement our own version of the publicly available likelihood \texttt{SimLow} \citepalias{2016A&A...596A.107P} and call it \simlow; differences in implementation will be highlighted below.}, \momento and \pydelfi. 

\subsection{Simulation-based likelihood} \label{sec:Likelihood_sim}
The simulation-based likelihood (\simbal), originally presented in \citetalias{2016A&A...596A.107P}, uses low-$\ell$ QCS estimates of only the $EE$ polarisation spectrum  to measure the optical depth. 

The joint sampling distribution for all the power spectrum elements is in general a function of all the power spectra components defining the model. However, the QCS procedure with reshaping, see Eq. \eqref{eq:cov_recast}, does a good job of approximately factorising this distribution by multipole.  Then the distribution of the power spectrum elements at a given multipole depends mainly on the theory elements at that multipole.  By only considering a single spectrum, $EE$ in this case, the requirement to handle intra-multipole correlations is avoided, motivating an approximate likelihood form: 
\begin{equation} \label{eq:likelihood}
    \mathcal{L}(\mat C|\mat{\Hat{C}}) = \prod_{\ell=\ell_{\rm min}}^{\ell_{\rm max}} \mathcal{L}_{\ell}(C_{\ell}|\Hat{C}_{\ell}) \ ,
\end{equation}
with $\ell_{\rm min}=2$ and $\ell_{\rm max}=29$, a product of one-dimensional functions. 

One then uses realisations generated according to a set of theory models to fit parametric forms to each of these one-dimensional sampling distributions.  This is done for each $\ell$ at the input theory $C_\ell$ values that happen to occur in the models considered. These fits are then evaluated at a realised $C_\ell$ value set equal to that of the real data $\Hat{C}_\ell$.  Finally, a further fit to these numbers, now as a function of the theory $C_l$, gives the effective likelihood function at that multipole. 

As the mask and noise do correlate the multipoles even with QCS, one may ask where such effects manifest themselves in the above procedure.  At a given $\ell=\ell_0$ say, one can imagine computing the marginalised one-dimensional sampling distribution for $\Hat{C}_{\ell_0}$ for a specific theory power spectrum.  Because of the couplings, in general this will differ between models even if they happen to share the same value of $C_{\ell_0}$.  The fitting procedure above then effectively averages over these distributions.  One expects the variations to be relatively modest for plausible models, and hopes that the theory models used to generate the realisations are close enough to reality not to lead to significant errors in the effective averaging. 

We therefore need to generate full maps of the observed CMB on which to measure the spectra. Considering the theory $C_{\ell}$'s for each mode over a large region of the parameter space would be computationally costly, since one would typically spend much time exploring low probability regions. We thus only explore the region of the power spectrum around the theory $C_{\ell}$ of interest. 

Our implementation of the simulation-based likelihood largely follows that of \citet{2020A&A...641A...5P}. The main difference is that we use Gaussian realisations of our noise fit to the \Planck simulations, rather than using the outputs of the noise simulations directly, given the limited number of the latter.  This allows us to use  many more independent noise realisations throughout the procedure. 

We generate 191 theoretical power spectra, $C_{\ell}(\tau, \theta)$, uniformly sampled over a range of $\tau$ values from $0.01-0.2$ inclusive with a step size of $\Delta \tau = 0.001$, where $\theta$ denotes all the cosmological model parameters defined in Sec.\ \ref{sec:posteriors}. Only $A_s$ is varied along with $\tau$, to keep the product $10^9A_s  e^{-2\tau}$ fixed at $1.870$, consistent with a high-$\ell$ likelihood constraint. For each of these theoretical power spectra, we generate 10000 Monte Carlo realisations of the CMB. Using the fitted NCM discussed in Sec. \ref{sec:NCM}, we generate Gaussian noise realisations that capture noise and systematics for the relevant frequency and detector set maps. At the map level, we combine the signal and noise maps. With this large suite of simulations we set up the simulation-based likelihood in the following way: 

\begin{enumerate}
    \item Compute low-$\ell$ QCS estimates of the simulations, $\Hat{C}_{\ell}^{\rm sim}$, from given theory power spectra that have values $C_{\ell}$.
    \item Compute the quasi-conditional $\mathcal{P}(\Hat{C}_{\ell}^{\rm sim} | C_{\ell})$ by fitting $\ell$-by-$\ell$ a model to the distribution of spectra. 
    We perform an unbinned
    maximum likelihood fit to the log-distribution, $\ln{\mathcal{P}(\Hat{C}_{\ell}^{\rm sim} |  C_{\ell})}$ as a function of $\Hat{C}_{\ell}^{\rm sim}$, using a third-order polynomial for the central part of the distribution and a first-order polynomial for its tails. Using Lagrange multipliers, we impose smoothness and continuity at the boundaries ensuring a good fit, $f_{\ell}(\Hat{C}_{\ell}^{\rm sim}| C_{\ell})$, to the conditional.
    \item We evaluate the above fits across the models at $\Hat{C}_{\ell}$, the QCS spectra for the data, fitting the same functional form as that used above but this time for the $C_{\ell}$ dependence.
\item Finally we construct our likelihood by combining the fits for each multipole as given in Eq.\ \eqref{eq:likelihood}, yielding
    \begin{equation} \label{eq:likelihood_slice}
        \ln \mathcal{P}(\mat{\Hat{C}}|\mat C) = \sum_{\ell = 2}^{29} \ln f_{\ell}(\Hat{C}_{\ell}| C_{\ell}) \ .
    \end{equation} 
\end{enumerate}
The method will be called \simbal throughout the paper and we shall denote our implementation of it by \simlow. In the present paper, we use \simlow with only polarisation data for the reason discussed above of the difficulties in handling correlations between variables.

\subsection{Likelihood approximation scheme} \label{sec:Likelihood_GLASS}
The General Likelihood Approximate Solution Scheme \citep[\glass
;][]{2017arXiv170808479G} was developed to allow a principled Bayesian
analysis of data even in situations where the sampling distribution
for the data is not fully known.  Our low-$\ell$ analysis of the CMB
polarisation is a case in point -- because of difficulties in
quantifying the noise in the maps, we choose to use quadratic
cross-spectra for robustness.  However, the joint distribution of the
multipoles of such spectra, computed on a masked sky, does not have a
simple analytic form.  Instead, \glass\ assumes one can compute
certain moments of functions of the data, here the spectral
multipoles, as a function of the parameters of the model under
investigation. One can then imagine \glass\ uses a maximum entropy
construction to compute a least-presumptive sampling distribution
consistent with these moments.  
We now introduce the method, briefly summarising \citet{2017arXiv170808479G}, to which the reader is referred for a fuller presentation. 

For an initial illustration, consider a situation in which one can obtain the mean $\Bar{x}(q)$ and variance $\sigma^2_x(q)$ of some function of the data $x$ in terms of a model parameter $q$, ideally analytically but also potentially via forward simulations.  Then, one maximises the entropy of the system
\begin{equation} \label{eq:entropy}
    H(\mathcal{P}) = - \int \mathrm{d}x\ \mathcal{P}(x) \ln{\frac{\mathcal{P}(x)}{\pi(x)}} \ 
\end{equation}
subject to the constraints on $\Bar{x}(q)$, $\sigma^2(q)$ and
normalisation of $\mathcal{P}(x)$, imposed via Lagrange multipliers.
The required values of the multipliers, $\lambda_x(q)$ and
$\lambda_{xx}(q)$, may be solved for, in general numerically, as a function of $q$.  The sampling distribution is then given as
\begin{equation}
    \label{eq:lagrange_multiplier}
    \mathcal{P}(x) = \frac{\pi(x) e^{-\lambda_xx - \lambda_{xx}x^2}}{\int dx \pi(x) e^{-\lambda_xx - \lambda_{xx}x^2}} = \frac{1}{\mathcal{Z}} \pi(x) e^{-\lambda_xx - \lambda_{xx}x^2}\ ,
\end{equation}
 where $\mathcal{Z}$ is a theory-dependent normalisation constant (also called the evidence or partition function).
 
Equivalently, this can be understood as solving for the \emph{action} (or negative log-likelihood) $\mathcal{S}$,
\begin{equation} \label{eq:action}
    \mathcal{S}(x, q) = -\ln{\pi(x)} + \lambda_x(q)x + \lambda_{xx}(q)x^2 + \ln{\mathcal{Z}(\lambda(q))} \ .
\end{equation}
The method naturally extends to multiple statistics $x^i,\ i \in \mathopen[1,\dots, n\mathclose]$ fitted to models with multiple parameters $q^{\alpha},\ \alpha \in \mathopen[1,\dots, m\mathclose]$, and using higher moments:
\begin{align} \label{eq:action_multi}
    \mathcal{S}(x, q) &= -\ln{\pi(x)} + \lambda_ix^i + \lambda_{ij}x^ix^j + \\ & \qquad + \lambda_{ijk} x^i x^j x^k + \dots + \ln{\mathcal{Z}(\lambda(q^{\alpha}))} \nonumber \ ,
\end{align}
where summation is implied over multiple indices. 

However, the evaluation of the Lagrange multipliers quickly becomes
very expensive, requiring the numerical computation of many
multi-dimensional integrals. 
We can avoid this cost by instead computing more moments.  To simplify notation, we
introduce a `meta-index' $I$ to first run over all indices $i$, then all pairs of indices $ij$, and so on.  $X^I$ then runs over the $x^i$, then the $x^i x^j$ and so on, and similarly $\lambda_I$ runs over the $\lambda_i$, then the $\lambda_{ij}$ and so on. 
We can express the moments of the $X^I$ as derivatives of the evidence $\mathcal{Z}$ with respect to the Lagrange multipliers:
\begin{align} 
    \langle X^I \rangle(\lambda) &= -\frac{\partial \ln \mathcal{Z}}{\partial \lambda_I} \ , \label{eq:GLASS_deriv1} \\
    \langle\langle X^IX^J \rangle\rangle(\lambda) &= \frac{\partial^2 \ln \mathcal{Z}}{\partial \lambda_I\partial \lambda_J} \label{eq:GLASS_deriv2}\ .
\end{align}
Next, we differentiate Eq.\ \eqref{eq:GLASS_deriv1} with respect to $q^{\alpha}$
\begin{equation}
    \langle X^I \rangle_{,a} = - \frac{\partial^2 \ln \mathcal{Z}}{\partial \lambda_J\partial \lambda_I} \frac{\partial \lambda_J }{\partial q^{\alpha}} = - \langle \langle X^IX^J \rangle \rangle\lambda_{J_{,a}}. \label{eq:GLASS_dxda}
\end{equation}
Now, differentiating the action in Eq.\ \eqref{eq:action_multi} yields
\begin{equation}
    \mathcal{S}_{,a} = (X^I - \langle X^I \rangle)\lambda_{J_{,a}} \ . \label{eq:sderiv}
\end{equation}
We can solve Eq.\ \eqref{eq:GLASS_dxda} for the $\lambda_{J_{,a}}$ in terms of the derivatives of the first moments and second order cumulants. Substituting into \eqref{eq:sderiv}, and adopting a vector/matrix notation to avoid explicitly writing meta-indices, we obtain 
\begin{equation}
    \mathcal{S}_{,a} = - (X - \langle X^\top \rangle)\langle \langle XX^\top \rangle \rangle \inv \langle X \rangle_{,a} \ ,\label{eq:dsda}
\end{equation}
which does not depend explicitly on the prior and the Lagrange multipliers. So, obtaining the moments by calculation or simulations, we can compute the gradient of $\mathcal{S}$.   This gradient can then subsequently be integrated between two points in parameter space in order to find the difference in approximate log-likelihood between the two models.

For an instructive if overly simple example, consider applying the above procedure in a 1-d problem in which the prior $\pi(x)$ is uniform and the first two moments of $x$ happen to be calculable as $\langle x \rangle = \mu$ and  $\langle x^2 \rangle = \mu^2 +\sigma^2$, 
where $\mu$ is a variable parameter of the model and $\sigma^2$ is fixed.  
Assume we will work to linear order in $x$ for $\mathcal{S}$. $I$ then
ranges over just one element, with  $X^0$ simply being $x$.  Knowing
up to second moments in $x$ is then sufficient to evaluate the single
component of $\langle\langle XX^\top \rangle\rangle$, which is $\langle x^2\rangle -\langle x \rangle^2=\sigma^2$. Eq. \eqref{eq:dsda} simply reads
\begin{equation}
    \mathcal{S}_{,\mu}= - (x - \mu) \frac{1}{\sigma^2} \cdot 1
\end{equation}
which in this case we can integrate by inspection to find $\mathcal{S}
= (x \mu - \mu^2 /2)/ \sigma^2$ up to a constant.  For inference of
$\mu$ this of course may be rewritten as $\mathcal{S} = (x - \mu )^2 /
(2 \sigma^2) + \mathrm{const}$, the exact Gaussian result that one might have anticipated from the form of the moments.  

For a more complicated example, consider observing a number of vectors
$\mat y$ of Gaussian-distributed components, with the vectors being
independent of each other but allowing components within a single vector to be correlated with each other according to a covariance matrix $\mathbf{C}$.  The sampling distribution is then
\begin{equation}
    \mathcal{P}(y|\mathbf{C})\diff^{2\ell+1}y =
    \frac{\diff^{2\ell+1}y}{|2\pi
    \mathbf{C}|^{(2\ell+1)/2}}e^{-\frac{1}{2}\sum_i y_i^{\top}
    \mathbf{C}\inv y_i} \ , \label{eq:cspectsampdiff}
\end{equation}
and we see that the components of the observed covariance matrix $\hat{\mathbf{C}}$,
\begin{equation}
    \hat{\mathbf{C}} \equiv\frac{1}{2\ell+1} \sum_i y_i y_i^{\top} \ ,
\end{equation}
serve as sufficient statistics for learning about the components of
$\mathbf{C}$.  The \glass\ scheme recovers the posterior associated
with \eqref{eq:cspectsampdiff} simply by working to linear order in
the components of $\hat{\mathbf{C}}$ for $\mathcal{S}$: after calculating their first and second moments, with some work Eq.\ \eqref{eq:dsda} recovers the optimal result
\begin{equation}
    \mathcal{S}_{\mathrm{true}} = \left(\ell+\frac{1}{2}\right)\left(  \tr \ \mathbf{C}\inv \hat{\mathbf{C}} + \ln{\frac{|\mathbf{C}|}{|\hat{\mathbf{C}}|}} -1\right) \ . 
\end{equation}

Hence we see how to use \glass to compute an
approximate likelihood, which we name \momento, for the cross-spectra
in our problem. One can in principle compute all of the intra- and
inter- $\ell$ cumulants between all the (\emph{TT}, \emph{TE} and)
\emph{EE} cross-spectra up to some required degree of approximation,
assuming the underlying maps are gaussianly-distributed (around some offset
noise template). In practice this is relatively easily manageable up to
fourth order in the spectra for moderate $\ell_\mathrm{max}$. Such moments are the
natural generalisations of the following formulae for a single multipole of
a single cross spectrum $\hat{C}_{12}$ on the full sky with isotropic
noise (and no unsubtracted map mean noise template): 
\begin{align}
\langle \hat{C}_{12} \rangle &= C \ ,  \label{eq:firstmom} \\
(2l+1) \langle\langle \hat{C}_{12}^2 \rangle\rangle &=
                                                      C^2+(C+N_{11})(C+N_{22})
                                                      \ , \label{eq:secondmom}\\ 
(2l+1)^2\langle\langle \hat{C}_{12}^3 \rangle\rangle &= 2C^3+6C(C+N_{11})(C+N_{22}) \ , \\
(2l+1)^3 \langle\langle \hat{C}_{12}^4 \rangle\rangle &= 6  \left( C^4+(C+N_{11})^2(C+N_{22})^2 \right. \nonumber \\
& \qquad \left. +\ 6 C^2 (C+N_{11})(C+N_{22}) \right) \ . \label{eq:lastmom}
\end{align}
Taking the theory power components themselves as parameters of the
theory, using such cumulants we can numerically integrate
$\mathcal{S}_{,a}$ up along a path from a fiducial model to a model in
question, for a selection of degrees of approximation (linear and
quadratic, requiring from quadratic up to fourth order moments of the
spectra). We find that even the linear approximation performs well and so 
typically use this in our work.

To summarise, \momento\ uses QCS power spectra, which use a
reasonable fiducial model with power spectrum $C_\ell^{\mathrm{fid}}$ to construct the
appropriate matrices.  Then, for each likelihood evaluation, \momento:
\begin{enumerate}
\item takes as input a set of theory $C_\ell$'s,
\item computes the difference $\Delta C_\ell \equiv
  C_\ell-C_\ell^{\mathrm{fid}}$ between the theory and the 
fiducial model
\item uses Romberg integration to compute the change in $\mathcal{S}$
going from the fiducial $C_\ell^{\mathrm{fid}} $'s to the theory
$C_\ell$'s along the line $C_\ell^{\mathrm{fid}}  + a \Delta
C_\ell$ in power spectrum space, with $a$ being a parameter
ranging from zero to one.  We choose to use a step size in $a$ of
$0.25$.
\item This requires the computation of gradients of $\mathcal{S}$ with respect to $a$ at four new positions
  in power spectrum space for every new likelihood evaluation (at
  $a=0.25, 0.5, 0.75, 1$), as
  those computed at the fiducial model ($a=0$) can be reused.
\item The gradients of $\mathcal{S}$  with respect to $a$ are linear combinations of
  those of $\mathcal{S}$  with respect to the associated $C_\ell$'s, and
\item these $\partial \mathcal{S}/ \partial C_\ell $'s  are 
computed via Eq. \eqref{eq:dsda}, which require both the QCS power spectra
of the data and
\item the multidimensional moments/cumulants of
  the QCS power spectra evaluated at the theory model corresponding to
  each $a$, via the appropriate multidimensional generalisations of
  Eqs. \eqref{eq:firstmom} - \eqref{eq:secondmom} in quick `linear'
  mode or Eqs. \eqref{eq:firstmom} - \eqref{eq:lastmom} in the fuller `quadratic' mode.
\end{enumerate}

In general, \glass is a very flexible scheme to compute principled posteriors where likelihoods are challenging to compute either for computational efficiency or more fundamental reasons. Since the approach is physically motivated, we do not have the `black box' behaviour seen in \delfi or other neural-network-based approaches. In contrast to the two other methods, which are dependent on the use of simulations to train their models, \momento\ only needs simulations for the computation of suitable NCMs, which are then used in the computation of moments as required.  

\subsection{Density-estimation likelihood-free inference} \label{sec:Likelihood_DELFI}

The final method takes an alternative approach to perform a simulation-based likelihood and is called the `likelihood-free' approach. In Sec.\ \ref{sec:Likelihood_sim} we fit a functional form to the likelihood $\mathcal{L}(C_{\ell}|\Hat{C}_{\ell})$ and obtained a total likelihood by assuming that each $\ell$ is independent. In this section we instead seek an invertible remapping of the measured cross-spectra, $\hat{C}_{\ell}$, to a set of variables $u_{\ell}$,  such that the resulting variables are statistically independent, Gaussian random variables. This mapping is obtained using a neural network (NN). With such a mapping the likelihood can be trivially evaluated as it is a multidimensional Gaussian combined with an appropriate Jacobian. The challenge with this approach is to obtain the mapping when the functional form of the distribution of the $\hat{C}_{\ell}$ is unknown\footnote{We remind the reader that for the \simbal approach an explicit functional form for the distribution of spectra was assumed.}.

More precisely, we start with a set of $N$ variables, $u_{\ell} $, that are unit-variance normal variables when conditioned on the parameters, \ie $u_{\ell}|\mathbf{\theta} \sim \mathcal{N}(0,I)$, where $\mathbf{\theta}$ denotes the conditional parameters (in our case $\tau$). We wish to find an invertible function such that $\hat{C}_{\ell}= f(u_{\ell})$. Given such a mapping we see that
\begin{align} \label{eq:refactorization}
\mathcal{P}(\hat{C}_{\ell} | \theta) \mathrm{d}^N\hat{C}_{\ell} & =   \mathcal{P}(  f(u_{\ell}) | \theta) \left | \frac{\dd \hat{C}_{\ell} }{\dd u_{\ell} } \right| \mathrm{d}^Nu_{\ell} 
\nonumber \\ & = \mathcal{P}_{u}(u_{\ell}|\theta)  \mathrm{d}^Nu_{\ell}  \ ,
\end{align}
where the density for $u_{\ell}$,\ $\mathcal{P}_{u}(u_{\ell}|\theta)$, is just a normal distribution. Finding the mapping $ f(u_{\ell})$, with a tractable Jacobian, seems a daunting task in general; the mapping from the $N$-dimensional normal  distribution to potentially multimodal distributions is likely non-trivial. To solve this challenge we use a second technique: the above remapping can be expressed as a series of $D$ simpler mappings, \ie
\begin{align}
\hat{C}_{\ell}= f(u_{\ell}) = f_D(f_{D-1}(\dots f_{1}(u_{\ell})))
\end{align}
with the only modification that Eq. \eqref{eq:refactorization} is changed to a product of Jacobians. The intuition is that we decompose the complex mapping into a series of simple transformations that slowly deform the probability density into a distribution that approaches that of the complex mapping\footnote{It has been shown that for some sufficiently flexible mappings arbitrary distributions can be modelled via such series of transformations \citep{2018arXiv180400779H, 2019arXiv190502325J}.}. We consider a particularly simple series of mappings. For the j$^{th}$ mapping we perform the following transformation:
\begin{align}
f_j^{i}(\mathbf{x} |\theta) = \frac{ \mat{x}^{i} - \mu^i_j(\mat{x}^{1...i-1},\theta)}{\sigma^i_j(\mat{x}^{1...i-1},\theta)} \ ,
\end{align}
\ie that the $i^{\mathrm{th}}$ output of the mapping is obtained by subtracting and scaling the $i^{\mathrm{th}}$ input by a function of all the previous ($i-1$) inputs. This has the nice property that the Jacobian in Eq. \eqref{eq:refactorization} for each transformation has the trivial form of the product of the functions $\sigma^i_j(x^{1...i-1},\theta)$  \footnote{We used the leading notation $\mu$ and $\sigma$ for these functions as this form can equally be thought of as stating that the distribution for the output $f^{i}(x^{i})$ follows a Gaussian distribution, conditioned on all the previous inputs.}. Thus, the desired distribution can be written as 
\begin{align}
\mathcal{P}(\hat{C}_{\ell} | \theta) \mathrm{d}^N\hat{C}_{\ell}  = \mathcal{P}(u_{\ell}|\theta) \prod\limits_{j=1}^{D} \prod \limits_{i=1}^{N} \sigma^i_j(u_\ell,\theta) \, \mathrm{d}^Nu_{\ell}\ .
\end{align} 
 This expression of the problem has shifted the complexity from fitting a functional form for the likelihood to identifying a suitable series of mapping functions $\sigma^i_j(u_\ell;\theta)$. 

To proceed we consider a family of functions  $\sigma^i_j(u_\ell,\theta | \mathbf{w}) $, parametrised by $\mathbf{w}$, and optimise the parameters to find the appropriate mapping. Equivalently stated, we solve a variational inference problem: we have a parametrised form for the likelihood $\mathcal{P}(\hat{C}_{\ell} | \theta; \mathbf{w})$ and we wish to optimise the parameters so that we can  approximate the true likelihood as accurately as possible. 

To find the best fitting weights of the NN, we minimise the Kullback-Leibler divergence $D_{\rm KL} (\mathcal{P}^*|\mathcal{P})$, between the parametric distribution $\mathcal{P}(\hat{C}_{\ell} | \theta; \mathbf{w})$ and the true distribution $\mathcal{P}^{*}(\hat{C}_{\ell}|\theta)$, which is defined as 
\begin{equation} \label{eq:KLdiv}
    D_{\rm KL} (\mathcal{P}^*|\mathcal{P})= - \int \mathcal{P}^*(\hat{C}_{\ell}|\theta) \ln \frac{\mathcal{P}(\hat{C}_{\ell}|\theta;\mathbf{w})}{\mathcal{P}^*(\hat{C}_{\ell}|\theta)} \mathrm{d}^N{\hat{C}}_{\ell} \ . 
\end{equation} 
The Kullback-Leibler divergence is a measure of the difference between probability distributions; it is a non-negative function that is zero only when the two distributions are identical. By minimising this function, we minimise the mismatch between our parametric conditional distribution and the true conditional distribution.

As we do not have access to the true distribution, only samples, we perform a Monte Carlo approximation of the Kullback-Leibler divergence using
\begin{align} \label{eq:loss_function}
    D^{\rm Monte\ Carlo}_{\rm KL} (\mathcal{P}^*|\mathcal{P}) = - \sum_{i=1}^{N_{\rm samples}} \ln{\mathcal{P}(\hat{C}_{\ell}|\theta;\mathbf{w}}),
\end{align}
which, as the number of samples tends to infinity, approaches Eq.\ \eqref{eq:KLdiv} up to an additive constant.

In principle, any family of functions can be used to perform the parametric fitting in this approach. However, if an overly restrictive set of functions is chosen we will have a poor approximation of the true distribution. Thus, in this work we choose the set of functions to be representable by a NN. NN are universal approximators, thus given sufficient data they can represent any function meaning we have a sufficiently flexible class of functions. 

We use the \pydelfi implementation of this method to construct a polarisation-only and a joint temperature-polarisation likelihood, with the precise configuration shown in App. \ref{sec:technical_details_delfi}. When dealing with a high-dimensional problem ($D\simgt 30$) in likelihood-free-inference, score compression is required both to reduce computational cost and give stable results. The greater the degree of compression, the more sub-optimal the likelihood. As a consequence, the results of our $TTTEEE$ \pydelfi likelihood have larger uncertainties, as discussed in detail in App. \ref{sec:delfi_compression} (for a more general discussion, see \cite{2018MNRAS.477.2874A}).

\section{Likelihood validation on simulations}
 \label{sec:compare_likelihoods}

\begin{figure}
    \centering
    \includegraphics[width=\columnwidth]{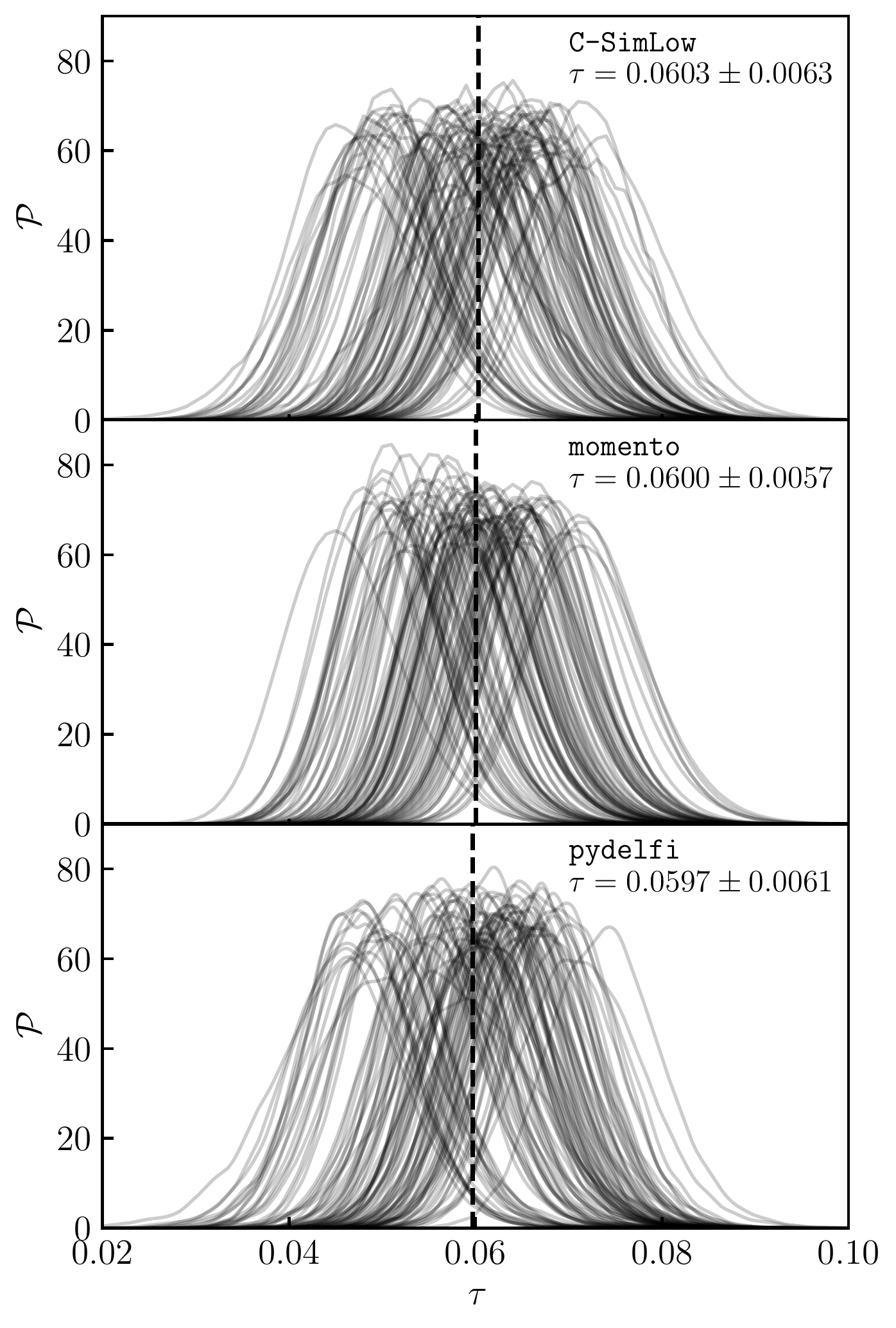}
    \vspace{-0.2in}
    \caption[comparison methods]{A test of the three likelihoods
      (\simlow, \momento and \pydelfi) on 100 simulated signal maps
      with an input $\tau=0.060$ and 100 noise and systematics maps
      from the \SRtwo end-to-end simulations. Each posterior per
      simulation is shown in black and the dashed black line is the
      mean of the maximum likelihood values for each method. The
      $\tau$ value in each top right corner shows the mean of the
      maximum likelihood values for $\tau$ over the simulations and the
      mean of the posterior widths.}
    \label{fig:likelihood_overfitting_posterior}
    \vspace{-0.1in}
\end{figure}

\begin{figure}
    \centering
    \includegraphics[width=\columnwidth]{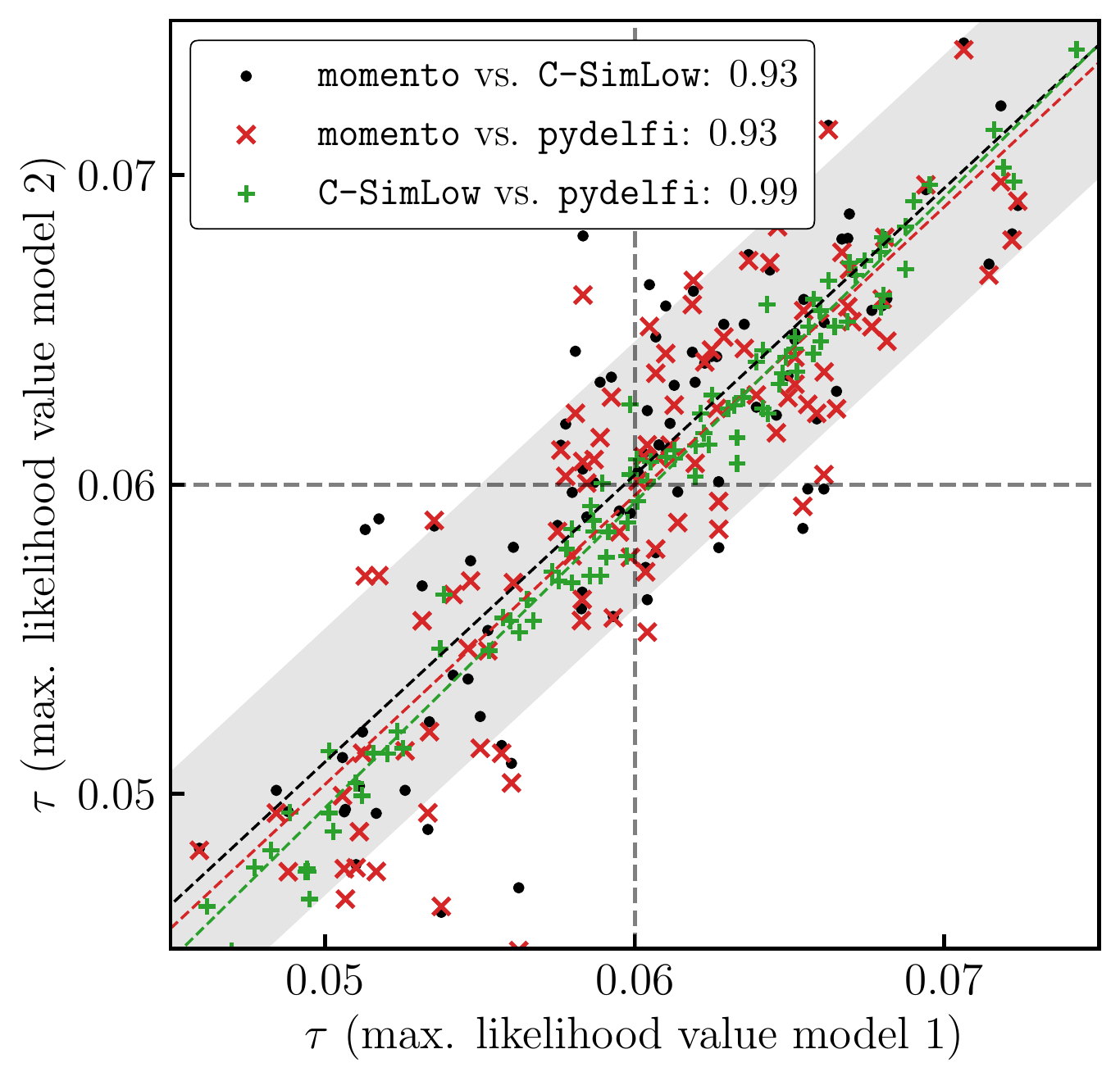}
    \vspace{-0.2in}
    \caption[correlation plot 3 methods]{Scatter plots between the maximum likelihood values corresponding to the posteriors shown in Fig. \ref{fig:likelihood_overfitting_posterior} for the three pairs of the likelihood methods (\simlow, \momento and \pydelfi). The grey shaded area shows the typical posterior width for the optical depth $\sigma(\tau)$. 
The black, red and green dashed lines show linear fits to each set of points. Their correlation coefficients (slopes of the linear fit) are given in the legend. There is a high degree of correlation between all three methods, especially between \simlow and \pydelfi.}
    \label{fig:likelihood_overfitting_posterior_scatter}
    \vspace{-0.1in}
\end{figure}

In this section we present tests of the three likelihood methods discussed in Sec. \ref{sec:Likelihood_sec}
 against simulations. The aim is  to investigate whether there are biases or significant differences in their 
performance.  We also test what happens if the input value of $\tau$ in the simulations is changed to be
higher or lower than the fiducial value of $0.06$ used to construct the covariance matrices required 
for the QCS estimator. 

Since end-to-end  simulations are used to model noise,  we have taken care to use an independent subset 
of the simulations for the tests described in this section. Thus, the noise estimation steps
used all but 100 end-to-end simulations for each \SRone and \SRtwo, leaving the remaining simulations 
available for likelihood validation. We also generated 100 new Gaussian CMB realisations 
for each of three values of $\tau$: $\tau = 0.05$, $0.06$ and $0.07$, using the same CMB realisations for
\SRone and \SRtwo. We use the same masks for the tests as for the data analysis (illustrated in Fig.\ \ref{fig:masks}).

Fig. \ref{fig:likelihood_overfitting_posterior} compares  the performance of the three likelihood methods for polarisation-only ($EE$) inferences on $\tau$ for the \SRtwo simulations (the \SRone case is very similar).  The figure
shows posteriors for each of the 100 test simulations using the $\tau=0.06$ CMB realisations. 
All three likelihoods perform satisfactorily, without significant bias.  The mean maximum likelihood values $\overline{\tau_{\rm ML}}$, mean of the posterior widths $\overline{\sigma(\tau)}$ and the standard deviation of the maximum likelihood values $\sigma(\tau_{\rm ML})$ found for each likelihood are given in  Table \ref{tab:overfitting_summary_result} (which also lists values
for the $\tau = 0.05$ and $0.07$ CMB realisations).

\begin{table*}\centering
{\def\arraystretch{1.2}\tabcolsep=10pt
\begin{tabular}{@{}l*{9}{c}@{}}\hline\hline
Likelihoods& \multicolumn{3}{c}{$\tau_{\rm in}=0.050$ }&\multicolumn{3}{c}{$\tau_{\rm in}=0.060$ }&\multicolumn{3}{c}{$\tau_{\rm in}=0.070$ }\\\hline
& $\overline{\tau_{\rm ML}}$      &$\overline{\sigma(\tau)}$ & $\sigma(\tau_{\rm ML})$& $\overline{\tau_{\rm ML}}$      &$\overline{\sigma(\tau)}$ & $\sigma(\tau_{\rm ML})$& $\overline{\tau_{\rm ML}}$      &$\overline{\sigma(\tau)}$ & $\sigma(\tau_{\rm ML})$\\\cmidrule(lr){2-4} \cmidrule(lr){5-7} \cmidrule(l){8-10} 
\simlow   & 0.0503 & 0.0064 & 0.0077 & 0.0603 & 0.0063 & 0.0069& 0.0703 & 0.0063 & 0.0066\\
\momento  & 0.0498 & 0.0056 & 0.0063 & 0.0600 & 0.0057 & 0.0063& 0.0704 & 0.0060 & 0.0065\\
\pydelfi  & 0.0496 & 0.0064 & 0.0077 & 0.0597 & 0.0060 & 0.0068& 0.0697 & 0.0058 & 0.0065\\
\hline
\end{tabular}}
\caption{Summary of likelihood tests performed using Gaussian realisations of CMB signal maps and 100 end-to-end \SRtwo simulations for the simulation-based likelihood (\simlow), the likelihood approximation scheme (\momento) and the density-estimation likelihood-free (\pydelfi) method. Here we make inferences on $\tau$ using low-$\ell$ polarisation data only. For each likelihood the mean maximum likelihood values $\overline{\tau_{\rm ML}}$, mean of the posterior widths $\overline{\sigma(\tau)}$ and the standard deviation of the maximum likelihood values $\sigma(\tau_{\rm ML})$ are computed. The input $\tau$ for the CMB realisations is denoted by $\tau_{\rm in}$. No evidence for any bias in the recovered $\tau_{\rm ML}$'s, even when the CMB is drawn from a distribution that does match the fiducial model with $\tau=0.06$, is seen.}
\label{tab:overfitting_summary_result}
\vspace{-0.1in}
\end{table*}

As well as investigating the average behaviour of the likelihoods, we have also compared them
realisation by  realisation. Fig. \ref{fig:likelihood_overfitting_posterior_scatter} shows scatter plots of the maximum likelihood values for each pair of  likelihoods for \SRtwo with an input $\tau$ of $0.06$. To guide the eye,
 the grey shaded area shows the range expected for a $\pm 1 \sigma$ error of $\delta \tau = 0.006$. 
There is a high degree of correlation between all three likelihoods, especially between \simlow and \pydelfi, as discussed in Sec. \ref{sec:Likelihood_DELFI}. This behaviour is expected since the \pydelfi approach is effectively a generalisation of the simulation-based likelihood \simlow (fitting a set of Gaussian's to the conditional distributions, instead of using a pre-defined functional form).  The scatter between  \simlow and \pydelfi is about a sixth of a sigma, and between either \simlow or \pydelfi and \momento it is roughly half a sigma\footnote{The quoted $\sigma$-shifts are calculated as a fraction of the scatter between the $\tau_{\rm ML}$ measurements, in other words of $\sigma(\tau_{\rm ML})$.}.  Applied to the same simulations, methodological differences in the 
likelihood implementation lead to differences in the maximum likelihood value of $\tau$ of less than a standard deviation.

Table \ref{tab:overfitting_summary_result} also gives results for simulations in which the CMB realisations are 
generated from models with both lower ($0.05$) and higher ($0.07)$ values of $\tau$ than the fiducial value $\tau = 0.06$ 
used to compute the QCS estimates. No bias is seen for any of the likelihoods confirming that the methods are
insensitive to the choice of fiducial cosmology.

We note that the posteriors on $\tau$ determined from \momento are about $10$\% tighter  than those determined from either 
\simlow or \pydelfi. The distribution of peak maximum likelihood values of $\tau$, shown in Fig. 
\ref{fig:likelihood_overfitting_posterior}, is also tighter for \momento. We have therefore  chosen to use
 \momento as our default low-$\ell$ likelihood in Sec. \ref{sec:posteriors} when combining with the high-$\ell$ $TTTEEE$ likelihood. Finally, all methods give average posterior widths that are slightly less than the scatter of their maximum likelihood values. The distribution of maximum likelihood values of $\tau$ 
 should be closely related to the width of the posterior distribution but is not guaranteed to be the same. The agreement is, however, close enough to demonstrate that the widths of the posterior distributions are not seriously in error.

\section{Results} \label{sec:posteriors}

In this section, we use  the three likelihoods described  in Sec.\ \ref{sec:Likelihood_sec} to 
derive constraints on $\tau$ from both the \Planck 2018 legacy maps (\SRone) and the 
\SRtwo maps.
In Sec. \ref{sec:results_100_143_constraints} we present results for the cross-correlation of the $100$ GHz and $143$ GHz full-mission maps, since this
choice of maps was used in \citetalias{Aghanim:2018eyx} and \citet{2019arXiv190809856P} to derive the results quoted in
 Eqs. \eqref{eq:Tau1} and \eqref{eq:Tau2}.
Sec. \ref{sec:results_sroll2_detset}  presents a more extensive investigation of the \SRtwo data set comparing all six combinations of $100$ GHz
and $143$ GHz  detector set maps to test whether the results are sensitive to different splits of the \Planck data.
To reduce  the computational burden, in Secs. \ref{sec:results_100_143_constraints} and \ref{sec:results_sroll2_detset} we perform one-dimensional parameter scans in $\tau$, 
allowing $A_s$ to change according to a fixed value of $10^9 A_s e^{-2 \tau} =1.870$. The  other 
parameters of the base $\Lambda$CDM cosmology fixed to $H_0 = 67.04$, $\Omega_{\rm b}h^2 = 0.0221$, $\Omega_{\rm c}h^2 = 0.12$, $\Omega_{\nu}h^2 = 0.00064$, $\theta_*=1.0411$, $n_{\rm s}=0.96$ \citepalias{Aghanim:2018eyx}. In Sec. \ref{sec:cosmo}, we relax the constraint on 
 $10^9 A_s e^{-2 \tau}$ and perform a  full Monte Carlo exploration of the six cosmological parameter space using \momento in conjunction with the high-$\ell$ \camspec v12.5HM likelihood \citepalias{EG19}.  

\subsection{Constraints using 100 $\times$ 143  full-mission QCS} \label{sec:results_100_143_constraints}

\begin{table}
\centering
{\def\arraystretch{1.1}\tabcolsep=5pt
\begin{tabular}{llll}
\hline \hline
Data Set     & Likelihood    & $\tau \ (EE)$      & $\tau \ (TTTEEE)$  \\ \hline
\Planck 2018 & \simlow  & $0.0530 \pm 0.0071$ & \multicolumn{1}{c}{\dots}\\ 
             & \momento & $0.0507 \pm 0.0063$ & $0.0527 \pm 0.0058$ \\ 
             & \pydelfi & $0.0517 \pm 0.0070$ & $0.0513 \pm 0.0078$\\ 
\SRtwo       & \simlow  & $0.0582 \pm 0.0057$ & \multicolumn{1}{c}{\dots}\\ 
             & \momento & $0.0581 \pm 0.0055$ & $0.0604 \pm 0.0052$\\ 
             & \pydelfi & $0.0588 \pm 0.0054$ & $0.0580 \pm 0.0064$\\ 
             \hline
\end{tabular}}
\caption{Summary of $\tau$ constraints for $100\times143$ full-mission cross-spectra obtained using a simulation-based likelihood (\simlow), a likelihood approximation scheme (\momento) and a density-estimation likelihood-free (\pydelfi) approach. For $\tau(EE)$ we measure $\tau$ only using the low multipole polarisation data and for $\tau(TTTEEE)$ we compute a joint likelihood for temperature and polarisation data. }
\label{tab:short_result_summary}
\vspace{-0.1in}
\end{table}

\begin{figure}
    \centering
    \includegraphics[width=\columnwidth]{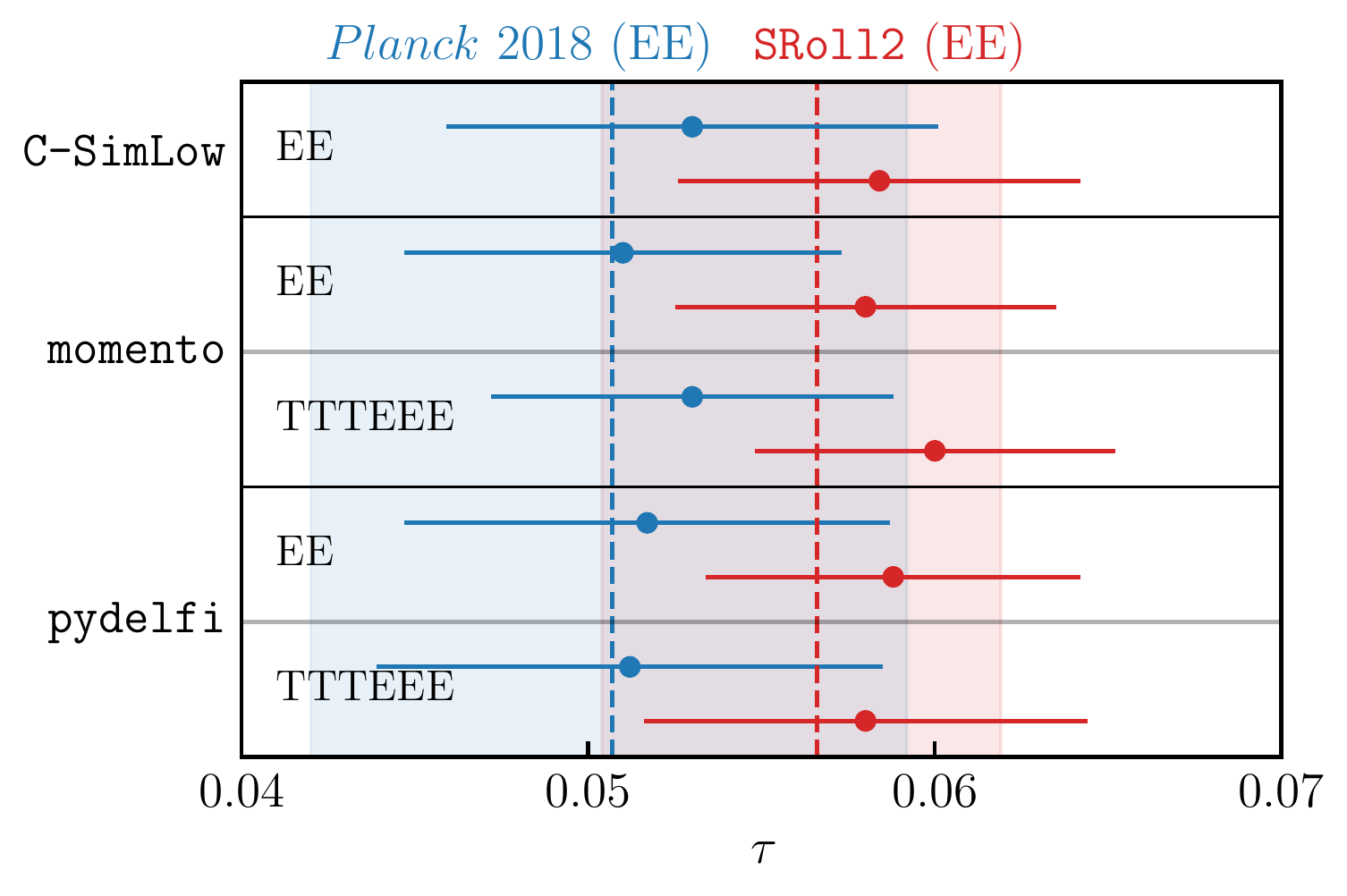}
    \vspace{-0.2in}
    \caption[Summary of posteriors.]{Summary plot of $\tau$ posteriors obtained from $100\times143$ full-mission \Planck 2018 and \SRtwo maps using a simulation-based likelihood (\simlow), a likelihood approximation scheme (\momento) and a likelihood-free inference (\pydelfi) approach. \pydelfi and \momento measure $\tau$ by, first, using only polarisation ($EE$) and, second, using temperature and polarisation data combined ($TTTEEE$). The vertical dashed lines and corresponding shaded regions indicate maximum likelihood values only using $EE$ polarisation data, for \Planck 2018 (\SRone) \citepalias{Aghanim:2018eyx} and \SRtwo \citep{2019arXiv190809856P}.}
    \label{fig:summary_posterior}
    \vspace{-0.1in}
\end{figure}

Table \ref{tab:short_result_summary} summarises the results from
foreground-cleaned $100\times143$ full-mission maps.  These results
are plotted in Fig.\ \ref{fig:summary_posterior} and compared with the
results of Eqs.\ \eqref{eq:Tau1} and \eqref{eq:Tau2} (shown as the
blue and red dashed lines respectively, together with $1\sigma$ errors
shown by the red and blue shaded areas). For all three likelihood
approximations, our results reproduce the upward movement in $\tau$
between \SRone and \SRtwo (as noted in
\cite{2019A&A...629A..38D}). Furthermore, the results from all three
likelihoods are consistent with each other. There are, however, some
interesting features that are worth noting:

\begin{table*}
\centering
{\def\arraystretch{1.3}\tabcolsep=4pt
\begin{tabular}{lccccccc}\hline\hline
  $\ \ \nu \times \nu'$ & \SRtwo        &  \simlow$ ($EE$){}^{*}$ &  \simlow  ($EE$)&  \momento ($EE$)       &  \momento ($TTTEEE$)       & \pydelfi ($EE$) & \pydelfi ($TTTEEE$)$^*$        \\
$\left[ \mathrm{GHz^2} \right]$&&${}^{*}$no synch.\ removal&&&&&${}^{*}$with compression\\\hline
100 $\times$ 143 & full $\times$ full & $0.0585 \pm 0.0057$ & $0.0582 \pm 0.0057$ & $0.0581 \pm 0.0055$ & $0.0604 \pm 0.0052$ & $0.0588\pm 0.0054$&$0.0580 \pm 0.0064$\\\cline{1-8}
\multirow{4}{*}{100 $\times$ 143} & ds1 $\times$ ds1 & $0.0642 \pm 0.0077$ & $0.0609 \pm 0.0076$ & $0.0595 \pm 0.0068$ & $0.0606 \pm 0.0066$ &$0.0594 \pm 0.0074$&$0.0603 \pm 0.0079$\\
& ds1 $\times$ ds2 & $0.0622 \pm 0.0073$ & $0.0589 \pm 0.0073$ & $0.0586 \pm 0.0067$ & $0.0603 \pm 0.0065$ &$0.0607 \pm 0.0072$&$0.0604 \pm 0.0076$ \\
& ds2 $\times$ ds1 & $0.0559 \pm 0.0071$ & $0.0538 \pm 0.0070$ & $0.0579 \pm 0.0067$ & $0.0589 \pm 0.0062$ &$0.0531 \pm 0.0066$ &$0.0526\pm 0.0081$\\
& ds2 $\times$ ds2 & $0.0597 \pm 0.0066$ & $0.0585 \pm 0.0068$ & $0.0570 \pm 0.0066$ & $0.0594 \pm 0.0061$ & $0.0579 \pm 0.0065$&$0.0575 \pm 0.0075$\\\cline{1-8} 
100 $\times$ 100 & ds1 $\times$ ds2 & $0.0719 \pm 0.0079$ & $0.0643 \pm 0.0079$ & $0.0638 \pm 0.0072$ & $0.0657 \pm 0.0071$ &$0.0619 \pm 0.0077$&$0.0662 \pm 0.0076$\\\cline{1-8} 
143 $\times$ 143 & ds1 $\times$ ds2 & $0.0702 \pm 0.0082$ & $0.0651 \pm 0.0073$ & $0.0676 \pm 0.0068$ & $0.0725 \pm 0.0067$ &$0.0656 \pm 0.0075$&$0.0703 \pm 0.0071$\\\cline{1-8}
mean & \dots & $0.0640 \pm 0.0075$ & $0.0603 \pm 0.0074$ & $0.0607 \pm 0.0068$ & $0.0629  \pm 0.0065$ &$0.0600  \pm 0.0072$ &$0.0612 \pm 0.0076$\\\hline
\end{tabular}}
\caption{Summary of $\tau$ posteriors derived from \SRtwo maps. Synchrotron and dust emission have been removed from the maps through template fitting, except for the column indicated which shows results obtained when only dust was subtracted. Hence the two \simlow columns illustrate the difference at the parameters level caused by the removal of synchrotron emission at the  map level. The bottom row assumes that the six detector based likelihoods are highly correlated and shows the mean of these results. The \pydelfi ($TTTEEE$) likelihood score compresses the power spectra before inferring the optical depth which results in wider posteriors than for the uncompressed \pydelfi ($EE$)-only likelihood.}
\label{tab:result_summary}
\vspace{-0.1in}
\end{table*}

\smallskip
\noindent
(i) The $TTTEEE$ \pydelfi results have larger error bars than the
$EE$ results alone, even though additional data is included in the
$TTTEEE$ likelihood (though the best-fit value of $\tau$ hardly
changes). As outlined in Sec. \ref{sec:Likelihood_DELFI} one of the drawbacks of likelihood-free inference is that higher dimensional problems ($D\simgt 30$) require additional data compression. In our $TTTEEE$ implementation we compressed the $84$ power spectrum
components for $2 \leq \ell \leq 29$ into three summary statistics,
one each for $TT$, $TE$ and $EE$, whereas for the $EE$-only
implementation we were able to avoid compression entirely, using all
29 $EE$ power spectrum multipoles.  The larger $TTTEEE$ \pydelfi error
is a consequence of lossy compression which actually degrades the
$EE$ block. In fact, we found that a score-compressed $EE$ posterior from \pydelfi gives a maximum likelihood value for $\tau$  that is  lower by
$\sim 0.25 \sigma$ compared to the results without compression. 
With compression the $TE$ data does actually pull the posterior upwards by $\sim 0.3 \sigma$, largely cancelling the effects of compression on $EE$. This is discussed further in App. \ref{sec:delfi_compression}. 

\smallskip

\noindent
(ii)  For \momento, adding $TT$ and $TE$ spectra to $EE$ causes 
upward shifts in $\tau$ of approximately $0.002$ ($\sim 0.4 \sigma$)
for both \SRone and \SRtwo $\tau$ values. The posterior width is reduced by a modest  $\sim 5\%$.
This behaviour is consistent with the parameter-shift criteria developed by 
\cite{2020MNRAS.499.3410G}. 

\smallskip

\noindent
(iii) The \SRtwo likelihoods consistently yield higher values of $\tau$, and with slightly tighter errors,
than the corresponding \SRone likelihoods. 

\smallskip

\noindent
(iv) The errors on $\tau$ from our application of \simlow to \SRone  are about $20\%$ smaller than those
 quoted in Eqs.\ \eqref{eq:Tau1}. There are two reasons for this: (a) the results of \citetalias{Aghanim:2018eyx}
used a sub-optimal noise model based on \Planck FFP8 end-to-end simulations for the QCS computations; (b) we subtracted a smoothed noise template at the map level (see Eq. \eqref{eq:smooth_template}), which reduces the size of the posterior widths by $\sim 5$--$10 \%$ for \SRone,  as  explained in more detail in App. \ref{sec:compare_SR20_SR10}.  

Recently, a new set of \Planck  maps for the LFI and HFI frequency bands have been developed \citep[][hereafter {\texttt NPIPE}] {2020arXiv200704997P}. As with \SRone and \SRtwo, a set of systematic
templates are fitted as part of the map-making stage. However, amongst other differences,  {\texttt NPIPE} retains
the CMB Solar dipole in each map and uses foreground polarisation priors at $30$, $217$ and $353$ GHz to break
parameter degeneracies. The use of polarisation priors leads to a suppression of the  polarisation signal at low multipoles,
necessitating the calibration of $EE$ power spectrum transfer functions from end-to-end numerical simulations. The transfer
functions corrections are quite large for the $EE$ multipoles $\ell = 2$-$7$ that 
contain most of the information on $\tau$. The analysis of the {\texttt NPIPE} $100\times 143$ $EE$  spectrum  presented in  \cite{2020arXiv200704997P} gives a value for $\tau$ that is lower by $1.2 \sigma - 1.6 \sigma$ compared to the results of  Table \ref{tab:short_result_summary}. The $\tau$ results from {\texttt NPIPE} are therefore broadly in agreement with those from \SRone and \SRtwo. 

\subsection{The optical depth from inter- and intra-frequency detector set combinations of \SRtwo maps} \label{sec:results_sroll2_detset}

To assess the robustness of the results of the previous section, we have analysed the
inter- and intra-frequency spectra computed from  \SRtwo detector set  maps. The \SRtwo cross-spectra used in this sub-section
are shown in Fig. \ref{fig:summary_QCS_cross_check}.

We stress that the likelihoods for each map pair have been computed/trained afresh using the appropriate
detector set noise covariance matrices constructed  from the relevant simulations.  The results of the cross-checks, for each of the three likelihoods,  are presented in Table \ref{tab:result_summary}. Each column lists the mean value for $\tau$ and the associated posterior width for the indicated spectrum combination. 

Focusing on the results from \momento, the
posterior widths for $143$ds1$\times143$ds2 are about $5\%$ smaller than for $100$ds1$\times 100$ds2, which is expected
because the  $143$ GHz maps are less noisy than the $100$ GHz maps. (The reduction in errors is, however, smaller in the other two likelihoods). More significantly, at multipoles $\ell=3-5$ the
$143$ds1$\times143$ds2 $EE$ spectrum lies above the $\tau=0.055$
theoretical line. As a consequence, for all likelihoods,  the
$143$ds1$\times143$ds2 value of $\tau$ is higher than that for the
$100\times 143$ spectra by about $1.4 - 2\sigma$.  The
$100$ds1$\times100$ds2 also shows a preference for higher values of
$\tau$ compared to the $100\times 143$ spectra, but to a lesser extent. This is 
suggestive of correlated residual systematics in the $100$ GHz detset 
and $143$ GHz detset maps which partially  cancel when $100$ GHz detset maps are 
cross-correlated against $143$ GHz detset maps 
\citep[as discussed by] [] {2019arXiv190809856P}.  The effects are relatively small, but in agreement 
with the conclusions of \cite{2019arXiv190809856P}, our results suggest that the $100\times 143$ spectra are likely to 
provide the most reliable constraints on $\tau$.

The last row in Table \ref{tab:result_summary} shows the mean of $\tau$ values for all of 
the inter- and intra-frequency cross detector 
set spectra constraints (ignoring correlations). The $100$ds1$\times 100$ds2 and $143$ds1$\times143$ds2 results  pull the 
means  to slightly higher values of $\tau$ compared to  the $100\times 143$ full-mission results, but only by about $0.5\sigma$.
Thus,  while there is some evidence of small systematic-related biases in 
  the $100$ds1$\times 100$ds2 and $143$ds1$\times143$ds2 $\tau$ values, the net effect of residual systematics on the $100 \times 143$ full-mission results are probably at the level of a standard deviation or less. This statement depends on the fidelity of the noise and systematics simulations.

Columns 3 and 4 of Table \ref{tab:result_summary} illustrate the impact of polarised synchrotron cleaning on $\tau$. (This test was done only for the \simlow likelihood). As expected, the effect on $\tau$ is most pronounced for $100$ds1$\times 100$ds2, with synchrotron cleaning lowering $\tau$ by about $1\sigma$ and bringing it into closer agreement with the $100 \times 143$ full-mission result. However, the effects of synchrotron cleaning on the $100 \times 143$ and $143$ds1$\times 143$ds2 spectra are significantly smaller, leading to changes in $\tau$ of $ \sim 0.3 \sigma$.  Synchrotron cleaning, while non-negligible, is not a critical factor in the $\tau$ constraints from the $100 \times 143$ spectra. 

\subsection{Full Monte Carlo Markov Chain Parameter Exploration} \label{sec:cosmo}

\begin{table}\centering
{\def\arraystretch{1.2}\tabcolsep=6pt
\begin{tabular}{@{}lcc@{}}\hline\hline
Likelihoods
& \makecell{\camspec ($TTTEEE$) + \\ \Planck low-$\ell$ $TT$+ \\ \momento ($EE$)}& \makecell{\camspec ($TTTEEE$) $+$ \\ \momento ($TTTEEE$) \\ \quad}\\\midrule
& \multicolumn{2}{c}{\SRone}\\\cline{2-3}
$\tau$       \dotfill                   &$0.0520^{+0.0055}_{-0.0062}$   & $0.0552^{+0.0056}_{-0.0065}$  \\
$\Omega_b h^2$       \dotfill           &$0.02226\pm 0.00014$           & $0.02228\pm 0.00015$          \\
$\Omega_c h^2$               \dotfill   &$0.1196\pm 0.0013$             & $0.1194\pm 0.0013$            \\
$100\theta_{\rm MC}$\dotfill            &$1.04103\pm 0.00028$           & $1.04108\pm 0.00028$          \\
$10^9 A_\mathrm{s} e^{-2\tau}$\dotfill  &$1.880\pm 0.011$               & $1.879\pm 0.012$              \\
$n_{\rm s}$     \dotfill                &$0.9669\pm 0.0043$             & $0.9677\pm 0.0044$            \\
$H_0$  \dotfill                         &$67.43\pm 0.55$                & $67.57^{+0.54}_{-0.60}$       \\
$\sigma_8$     \dotfill                 &$0.8086\pm 0.0066$             & $0.8101\pm 0.0066$            \\
$z_{\rm re}$           \dotfill         &$7.45\pm 0.62$                 & $7.77\pm 0.61$                \\
& \multicolumn{2}{c}{\SRtwo }\\\cline{2-3}
$\tau$ \dotfill                         &$0.0592^{+0.0051}_{-0.0058}$   & $0.0627^{+0.0050}_{-0.0058}$  \\
$\Omega_b h^2$   \dotfill               &$0.02229\pm 0.00015$           & $0.02231\pm 0.00014$          \\
$\Omega_c h^2$   \dotfill               &$0.1194\pm 0.0013$             & $0.1191\pm 0.0013$            \\
$100\theta_{\rm MC}$  \dotfill          &$1.04107\pm 0.00028$           & $1.04109\pm 0.00027$          \\
$10^9 A_\mathrm{s} e^{-2\tau}$ \dotfill &$1.879\pm 0.011$               & $1.878\pm 0.011$              \\
$n_{\rm s}$      \dotfill               &$0.9678\pm 0.0044$             & $0.9688\pm 0.0047$            \\
$H_0$       \dotfill                    &$67.56\pm 0.57$                & $67.67\pm 0.57$               \\
$\sigma_8$          \dotfill            &$0.8136\pm 0.0063$             & $0.8155\pm 0.0062$            \\
$z_{\rm re}$    \dotfill                &$8.18\pm 0.54$                 & $8.51\pm 0.52$                \\
\hline
\end{tabular}}
\caption{Cosmological parameter constraints for $\Lambda$CDM cosmology obtained using \momento with $100\times143$ full-mission spectra at low-$\ell$ and the \camspec v12.5HM ($TTTEEE$) at high-$\ell$  with 68\% confidence levels. We compare the effect on the parameter constraints by combining our polarisation-only or joint temperature-polarisation low-$\ell$ likelihood \momento with the high-$\ell$ ($\ell\geq 30$) likelihood \camspec. The redshift of reionization $z_{\rm re}$ is defined in the same way as in \citetalias{Aghanim:2018eyx}.}
\label{tab:cosmo_params}
\vspace{-0.1in}
\end{table}

We explore the full $\Lambda$CDM cosmological parameter space by combining our full-mission $100\times143$  \momento  $EE$ and $TTTEEE$ likelihoods
at low multipoles 
for both \SRone and \SRtwo with  the high-$\ell$ \camspec v12.5HM ($TTTEEE$) likelihood (which uses \SRone maps).
The  $EE$ \momento likelihood uses the multipole range 
$2\leq \ell \leq 29$ and is utilised with the \Planck 2018 low-$\ell$ $TT$ likelihood over
the same multipole range. The \momento $TTTEEE$ likelihood uses the  multipole range $2\leq \ell \leq 10$ to speed up the
low multipole likelihood evaluations. The multipoles $11-29$ in $TE$ and $EE$ have very little constraining power on $\tau$\footnote{We tested this by constraining $\tau$ on simulations for the
  reduced multipole range ($2\leq \ell \leq 10$) compared to the full low
  multipole range ($2 \leq \ell \leq 29$) and found that the
  posteriors were almost identical.} and so little information on
$\tau$ is lost by truncating the \momento\ likelihood at $\ell =
10$. The \momento $TTTEEE$ likelihood is used with the
\Planck 2018 low-$\ell$ $TT$ likelihood over the multipole range
$11\leq \ell \leq 29$, so that there are no multipole gaps in the $TT$
likelihood.

Table \ref{tab:cosmo_params} lists the results of full MCMC exploration of the parameters of the base $\Lambda$CDM  cosmology.
 These results are  similar to those summarised in Table \ref{tab:short_result_summary} for the 
one-dimensional $\tau$ scans. The \SRtwo results for $\tau$ are about $1\sigma$ higher than those from
\SRone,  and the $TTTEEE$  \momento likelihoods give values for $\tau$ that are about  $0.3\sigma$ higher than 
those using the  $EE$ \momento likelihoods.  The  values for $10^9 A_s e^{-2 \tau}$ are also within about $1 \sigma$ 
of the value assumed for the one-dimensional $\tau$ scans.  

We can compare the results from \momento\ with those of full parameter
analyses using the \simbal\ $EE$ likelihood combined with the low multipole $TT$ likelihood and the $\texttt{Plik}$
high multipole $TTTEEE$ likelihood reported in
\citetalias{Aghanim:2018eyx} and \citet{2019arXiv190809856P}:
\begin{subequations} 
\begin{align}
\quad &\tau = 0.0544^{+0.0070}_{-0.0081},  \qquad  \SRone, \label{eq:tau_TTTEEE_cosmo1}\\ 
\quad &\tau = 0.0591^{+0.0054}_{-0.0068},  \qquad \SRtwo.  \label{eq:tau_TTTEEE_cosmo2}
\end{align}
\end{subequations}
Our results using \momento are higher by about $0.3 - 0.5 \sigma$ and (formally) have slightly smaller error bars.

\begin{figure}
    \centering
    \includegraphics[width=\columnwidth]{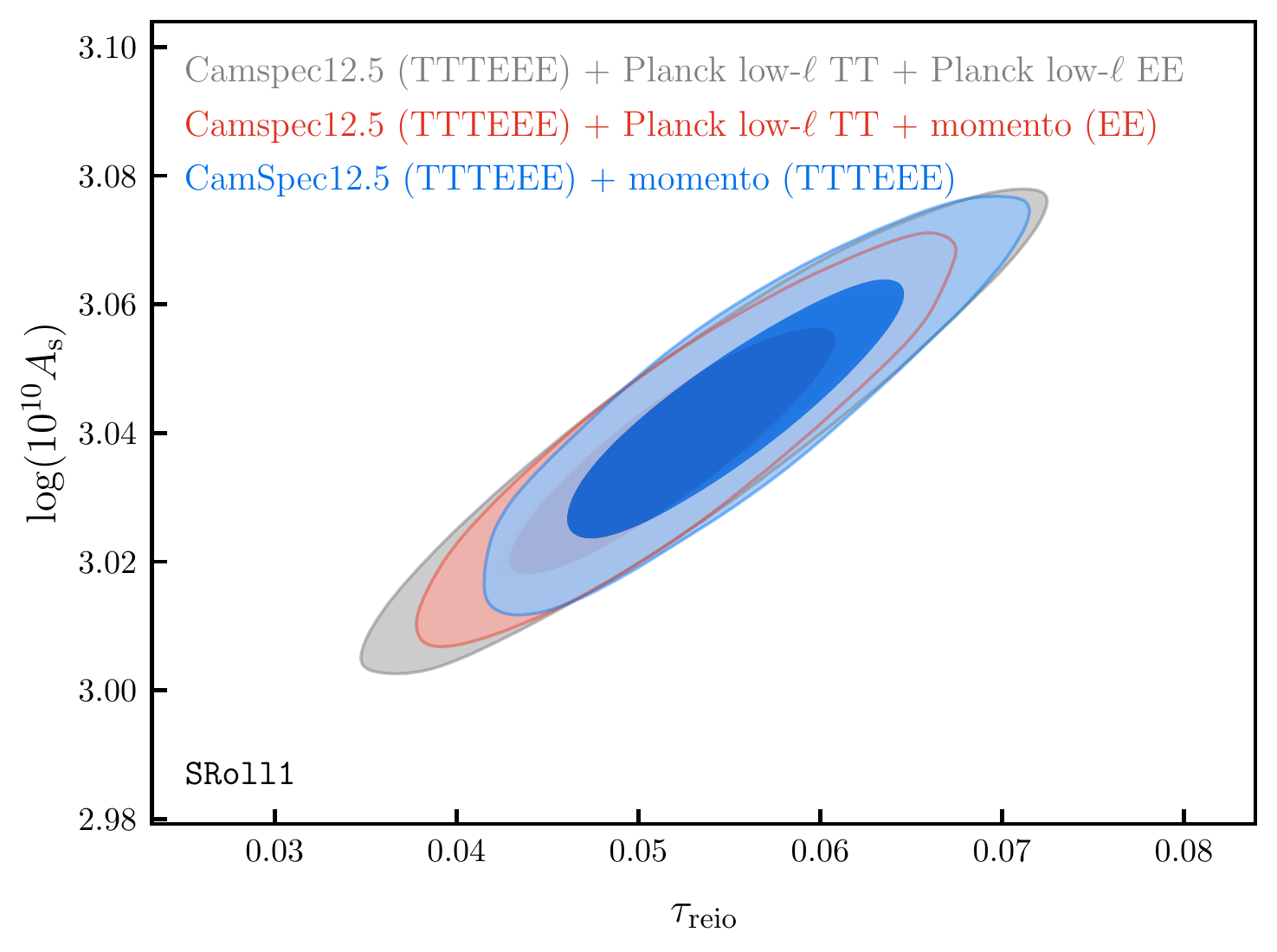}
    \vspace{-0.2in}
    \caption[parameter constraints]{Parameter constraints for the base $\Lambda$CDM cosmology for $\tau$ and $\ln(10^{10}A_s)$. The high-$\ell$ likelihood is \camspec v12.5HM ($TTTEEE$) and the three different low-$\ell$ likelihood combinations are: (i) \Planck $TT$ together with \Planck $EE$ (grey), (ii) \Planck $TT$ together with \momento ($EE$) (red) and (iii) \momento ($TTTEEE$) with \Planck $TT$ (restricted to $11 \leq \ell \leq 29$) (blue).}
    \label{fig:param_constraint}
    \vspace{-0.1in}
\end{figure}

\begin{figure}
    \centering
    \includegraphics[width=\columnwidth]{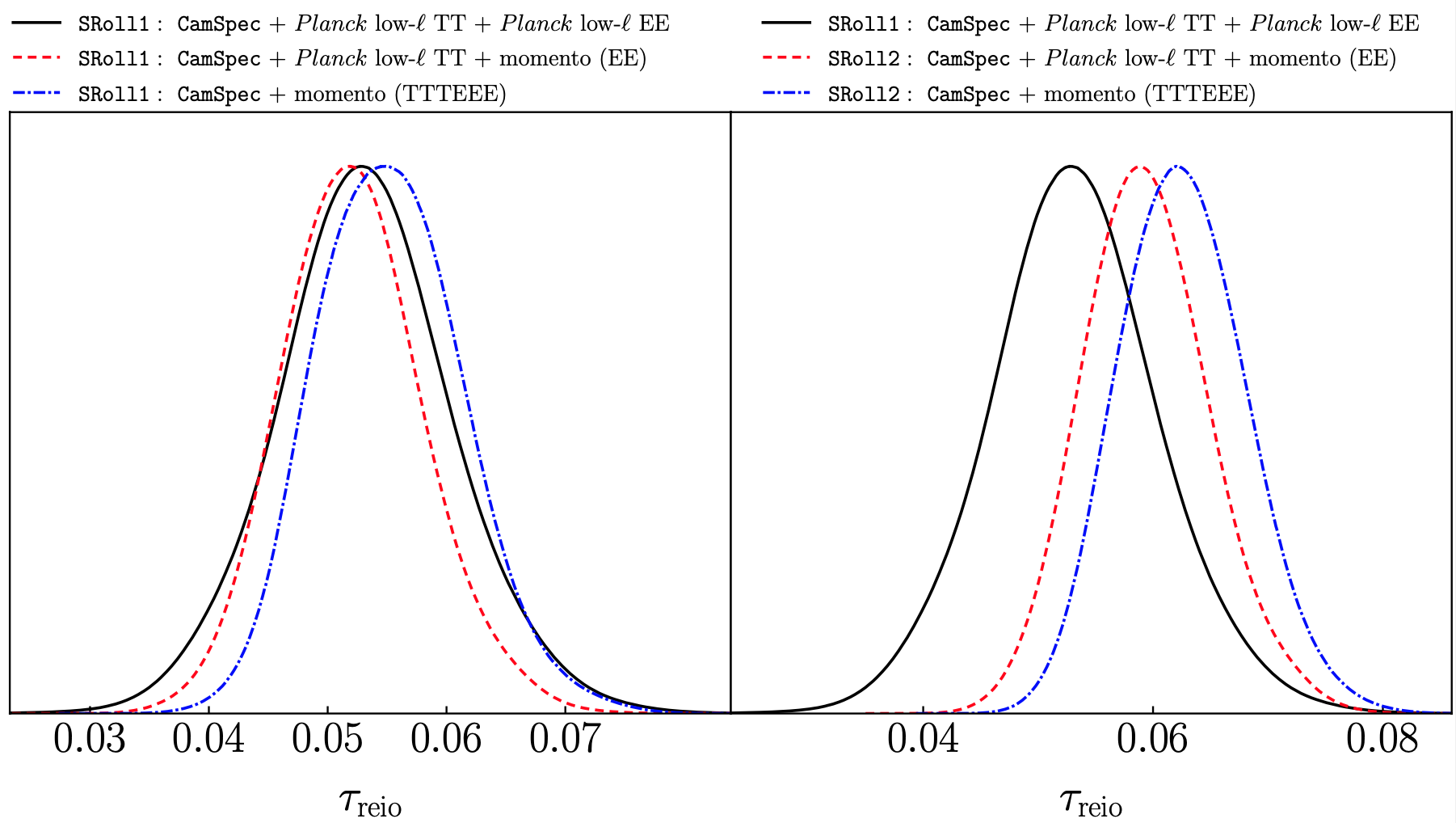}
    \vspace{-0.2in}
    \caption[Summary of posteriors Planck 2018.]{Marginalised posterior distributions for $\tau$ derived from the
 high-$\ell$ \camspec v12.5HM ($TTTEEE$) likelihood combined with low-$\ell$ likelihoods as discussed in the
text. The black lines in both plots use the low-$\ell$ \Planck 2018 $\texttt{SimLow}$ polarisation likelihood as in \citetalias{Aghanim:2018eyx}.
The results of replacing $\texttt{SimLow}$ with \SRone and \SRtwo  \momento likelihoods are shown in the left and right hand
plots, respectively.}
    \label{fig:marginalized_posterior_camspec_momento}
    \vspace{-0.1in}
\end{figure}

Fig. \ref{fig:param_constraint} illustrates the changes to the $\tau$ constraints caused by switching from $\texttt{SimLow}$ to \momento $EE$ and then to \momento $TTTEEE$. This figure  shows contours in the $\ln 10^{10}A_s$--$\tau$ plane for base $\Lambda$CDM. In each case
we use the same high-$\ell$ \camspec v12.5HM $TTTEEE$ likelihood for $30 \leq \ell \leq 2500$
 and so only the low-$\ell$ likelihoods change:  grey  contours for $\texttt{SimLow}$, red contours for \SRone \momento $EE$ and
blue for \SRone \momento $TTTEEE$. Interestingly, the \momento $TTTEEE$ likelihood disfavours values of
$\tau \simlt 0.04$ that are already excluded by the Gunn-Peterson test (see Eq. \eqref{eq:tau_eq}). In other words, the
posteriors of the \momento likelihood are consistent with what we know about the intergalactic medium.

The main results of this section are summarised in Fig. \ref{fig:marginalized_posterior_camspec_momento}. The plot to the
left shows the marginalised posterior distribution for $\tau$ for \camspec combined with the \SRone likelihoods as described 
above. This shows the small shifts in $\tau$ when we use the \SRone \momento likelihoods 
in place of the \Planck\ 2018 low-$\ell$ $EE$  likelihood. The right hand plot shows how the posteriors change if we use the 
\SRtwo \momento likelihoods. The \momento constraints shift to higher values of $\tau$ as a consequence of the changes to
the HFI 
map-making. We take as our `best' estimate of $\tau$ and redshift of reionization, $z_{\rm re}$, the results from
the combined \camspec+ \SRtwo \momento $TTTEEE$ likelihood:
\begin{equation}
\tau =  0.0627^{+0.0050}_{-0.0058},  \qquad z_{\rm re} = 8.51 \pm 0.52,   \label{eq:rogers_final_answer}
\end{equation}
slightly higher than the result of Eq. \eqref{eq:tau_TTTEEE_cosmo2}.

\section{Summary and conclusions} \label{sec:Conclusions}

The determination of the optical depth to reionization $\tau$ from the
CMB is extremely challenging, yet of great importance for our
understanding of the intergalactic medium and the formation of the
first stars and galaxies. Improvements in the HFI map-making
algorithms described in \citetalias{2016A&A...596A.107P} and \citetalias{2019A&A...629A..38D} have led to  maps which
have low levels of residual systematics in polarisation resulting in 
low values of $\tau$ (Eqs.
\eqref{eq:tau_TTTEEE_cosmo1} and \eqref{eq:tau_TTTEEE_cosmo2}). Producing 
high fidelity polarisation maps is only one part of the story however.
To derive accurate constraints on $\tau$ requires an accurate likelihood.
The construction of a $TTTEEE$ likelihood  is not straightforward at low
multipoles for maps with complex noise properties and partial sky coverage.
There is no analytic guide to help create such a likelihood, particularly
if the likelihood is built around quadratic cross-spectra.  In \citetalias{2016A&A...596A.107P} and 
\citetalias{2019A&A...629A..38D}, an $EE$ likelihood was constructed based on a relatively
small number of end-to-end simulations, leading to the results of Eqs.
\eqref{eq:tau_TTTEEE_cosmo1} and \eqref{eq:tau_TTTEEE_cosmo2}.

In this paper we have developed and compared  three likelihood techniques on the \Planck
\SRone and \SRtwo maps. The first is a variant of the \simbal scheme described in 
\citetalias{2016A&A...596A.107P} and \citetalias{2019A&A...629A..38D}, but using more accurate simulation-based noise covariance matrices
to construct quadratic cross-spectra and to generate a large number of independent
noise realisations. The second (\momento) is based on the \glass 
maximum entropy approach developed by \cite{2017arXiv170808479G} and the third 
(\pydelfi) is a density-estimation `likelihood free' scheme that follows closely the implementation 
described by \cite{2019MNRAS.488.4440A}. The \momento and \pydelfi approaches can
be generalised to construct low multipole $TTTEEE$ likelihoods. (Though not explored
in this paper, it is straightforward to adapt these schemes to develop likelihoods
incorporating other low multipole spectra, e.g. $BB$, $ET$.)

Our main conclusion is that all three likelihood methods are in good agreement and support the conclusions
on $\tau$ reported in  \citetalias{Aghanim:2018eyx} and \citet{2019arXiv190809856P}; we do, however, see small differences
between the likelihoods as summarised in Tables \ref{tab:short_result_summary} and \ref{tab:result_summary}. Using only the
spectra at low multipoles, our results tend to give higher values of $\tau$ than those using \simbal (Eqs. \eqref{eq:Tau1}) and 
\eqref{eq:Tau2})  by up to $\sim 0.8 \sigma$. However, if we include  the high multipole \camspec  $TTTEEE$ likelihood, the results for $\tau$ using the \SRtwo \momento $EE$ likelihood is  very close to that
given in Eq. \eqref{eq:tau_TTTEEE_cosmo2} though with a smaller formal error.

We constructed low multipole $TTTEEE$ likelihoods using \momento and
\pydelfi. For \momento, using a $TTTEEE$ likelihood leads to  smaller
errors on $\tau$ than using $EE$ alone, as expected. However, we had to apply 
data compression to produce a \pydelfi  $TTTEEE$ likelihood that was numerically fast and robust enough 
for likelihood analysis. This resulted in a loss of information and to $\tau$
constraints  that had slightly larger errors using  \pydelfi $TTTEEE$ compared
to using \pydelfi $EE$, though with no evidence of any bias; see App. \ref{sec:delfi_compression} for a detailed discussion.

We also made a thorough analysis of different detector set data splits at $100$ and $143$ GHz, as summarised in Table 
\ref{tab:result_summary}. For all likelihoods, the $100$ds1$\times100$ds2 and $143$ds1$\times143$ds2 spectra
give $\tau$ values that are higher than those from the baseline $100 \times 143$ full-mission analysis by between
$0.8$ and $2 \sigma$. There is therefore evidence that within a frequency band  there remain correlated systematic
effects (as is apparent visually from Fig. \ref{fig:cleaned_maps2}) that bias $\tau$ high by $\sim 0.01$. The series
of null tests described in \citetalias{2019A&A...629A..38D}, together with the absence of any statistically significant 
$B$-mode signal at low multipoles, suggests that the
$100 \times 143$ full-mission cross-spectra should provide unbiased estimates of $\tau$. The changes  between  \citetalias{2016A&A...596A.107P} and \citetalias{2019A&A...629A..38D} suggest an upper bound of about $1 \sigma$ to  biases in $\tau$ caused by residual systematics.

As noted above, the likelihood techniques explored here have wider applications and can be adapted to other problems involving
low multipole polarisation maps, particularly if the maps have complex noise properties. An obvious example is
the measurement of the tensor-to-scalar ratio $r$, in addition to $\tau$, from the forthcoming CMB satellite 
\emph{LiteBIRD} \citep{2020JLTP..199.1107S}.

\section*{Acknowledgements}
RdB is grateful to Pablo Lemos, Oliver Friedrich and Will Handley for fruitful discussions regarding likelihood-free inference, to Erik Rosenberg for help with \camspec and to Paul Murdin for suggestions on an early version of the manuscript. We are indebted to the members of the \Planck collaboration for their enormous efforts in producing such a wonderful set of data and especially thank the Bware team for their \SRtwo products. 

RdB acknowledges support from the Isaac Newton Studentship, Science and Technology Facilities Council (STFC) and Wolfson College, Cambridge. SG acknowledges the award of a Kavli Institute Fellowship at KICC. W.R.C. acknowledges support from the UK Science and Technology Facilities Council (grant number ST/N000927/1), the World Premier International Research Center Initiative (WPI), MEXT, Japan and the Center for Computational Astrophysics of the Flatiron Institute, New York City. The Flatiron Institute is supported by the Simons Foundation.

We acknowledge the use of: \texttt{PolyChord} \citep{2015MNRAS.450L..61H, 2015MNRAS.453.4384H}, \texttt{HEALPix} \citep{2005ApJ...622..759G}, \texttt{CAMB} (\url{http://camb.info}), \texttt{GetDist} (\url{https://getdist.readthedocs.io}), \texttt{Cobaya} \citep{2020arXiv200505290T}, \pydelfi (\url{https://github.com/justinalsing/pydelfi}), \texttt{ChainConsumer} \citep{Hinton2016} and \texttt{emcee} \citep{emcee}.

This work was performed using the Cambridge Service for Data Driven Discovery (CSD3), part of which is operated by the University of Cambridge Research Computing on behalf of the STFC DiRAC HPC Facility (\url{https://dirac.ac.uk}). The DiRAC component of CSD3 was funded by BEIS capital funding via STFC capital grants ST/P002307/1 and ST/R002452/1 and STFC operations grant ST/R00689X/1. DiRAC is part of the National e-Infrastructure.

\section*{Data availability}
The data underlying this article will be shared on reasonable request. The \Planck 2018 (\SRone) and \SRtwo frequency maps and end-to-end simulations are publicly available: \SRone simulations (labelled FFP10) at \url{https://pla.esac.esa.int}; \SRtwo simulations at \url{http://sroll20.ias.u-psud.fr}. 



\bibliographystyle{mnras}
\bibliography{references} 


\appendix

\section{Analytic solution to the noise inference problem}\label{sec:analytic_solution}

In order to minimise Eq. \eqref{eq:sfornoise} analytically for $\mat \Psi$, we first consider the variation in $\delta \mathcal{S}$ caused by small changes in $\mat M$. Omitting the $n_s/2$ prefactor, we have:
\begin{align} 
    \delta \mathcal{S} &= \tr(\delta (\mat M \inv) \,  \Hat{\mat N}) + \delta \ln |\mat M| \\
    &= \tr \left[ -\mat M\inv (\delta \mat M) \mat M \inv \Hat{\mat N} + \mat M \inv (\delta \mat M) \right] \\
    &= \tr\left[ -\mat M\inv \mat Y (\delta \mat{\Psi})\mat Y^\top \mat M \inv \Hat{\mat N} + \mat M \inv \mat Y (\delta \mat{\Psi})\mat Y^\top  \right]
\end{align}
using the standard results for matrices that $\delta (\mat M \inv ) = - \mat M \inv (\delta \mat M) \mat M \inv $ and $\delta \ln |\mat M | = \tr \mat M \inv \delta \mat M $, and $\delta \mat M = \mat Y \delta \mat \Psi \mat Y^\top $.  
Next we use the cyclic property of the trace to obtain:
\begin{align}
    \delta \mathcal{S} 
    &= \tr \left[ \delta \mat{\Psi} \left( \mat Y^\top \mat M \inv (  \mat M -  \Hat{\mat N})  \mat M \inv \mat Y  \right) \right] . 
\end{align}
At the minimum we require $\delta \mathcal{S}=0$ for arbitrary $\delta \Psi$ and thus need
\begin{align}
\mat Y^\top \mat M \inv (  \mat M -  \Hat{\mat N})  \mat M \inv \mat Y  = \mat 0.    \label{eq:spartway}
\end{align}
We now use the generalised Sherman–Morrison–Woodbury formula
\begin{equation} \label{eq:SMW_eq1}
    \mat M\inv = \mat N \inv - \mat N \inv \mat Y(\mat{\Psi}\inv +\mat Y^\top \mat N \inv \mat Y)\inv \mat Y^\top \mat N \inv 
\end{equation}
(and adding and subtracting $\mat \Psi \inv $) to rewrite $\mat Y^\top \mat M\inv$ as
\begin{align} 
    \mat Y^\top \mat M\inv &= \mat Y^\top \left( \mat N\inv - \mat N\inv \mat Y(\mat{\Psi}\inv +\mat Y^\top \mat N\inv \mat Y)\inv \mat Y^\top \mat N\inv  \right)\\
    &= \mat Y^\top \mat N\inv - (\mat Y^\top \mat N\inv \mat Y + \mat{\Psi}\inv - \mat{\Psi}\inv) \nonumber \\
    & \qquad \qquad \times (\mat{\Psi}\inv + \mat Y^\top \mat N\inv \mat Y)\inv \mat Y^\top \mat N\inv \\
    &= \mat{\Psi}\inv (\mat{\Psi}\inv + \mat Y^\top \mat N\inv \mat Y)\inv \mat Y^\top \mat N\inv \ .
\end{align}
Substituting this and its transpose into Eq.\ (\ref{eq:spartway}) yields
\begin{align}
    &\mat Y^\top \mat N\inv \left[ \mat N + \mat Y\mat{\Psi} \mat Y^\top -\Hat{\mat N} \right] \mat N\inv \mat Y = \mat 0.
\end{align}
Rearranging this for $\mat \Psi$ then gives the desired Eq.\ (\ref{eq:solnforpsi}). 

\section{Smoothed Template Subtraction}\label{sec:compare_SR20_SR10}
\begin{figure}
    \centering
    \includegraphics[width=\columnwidth]{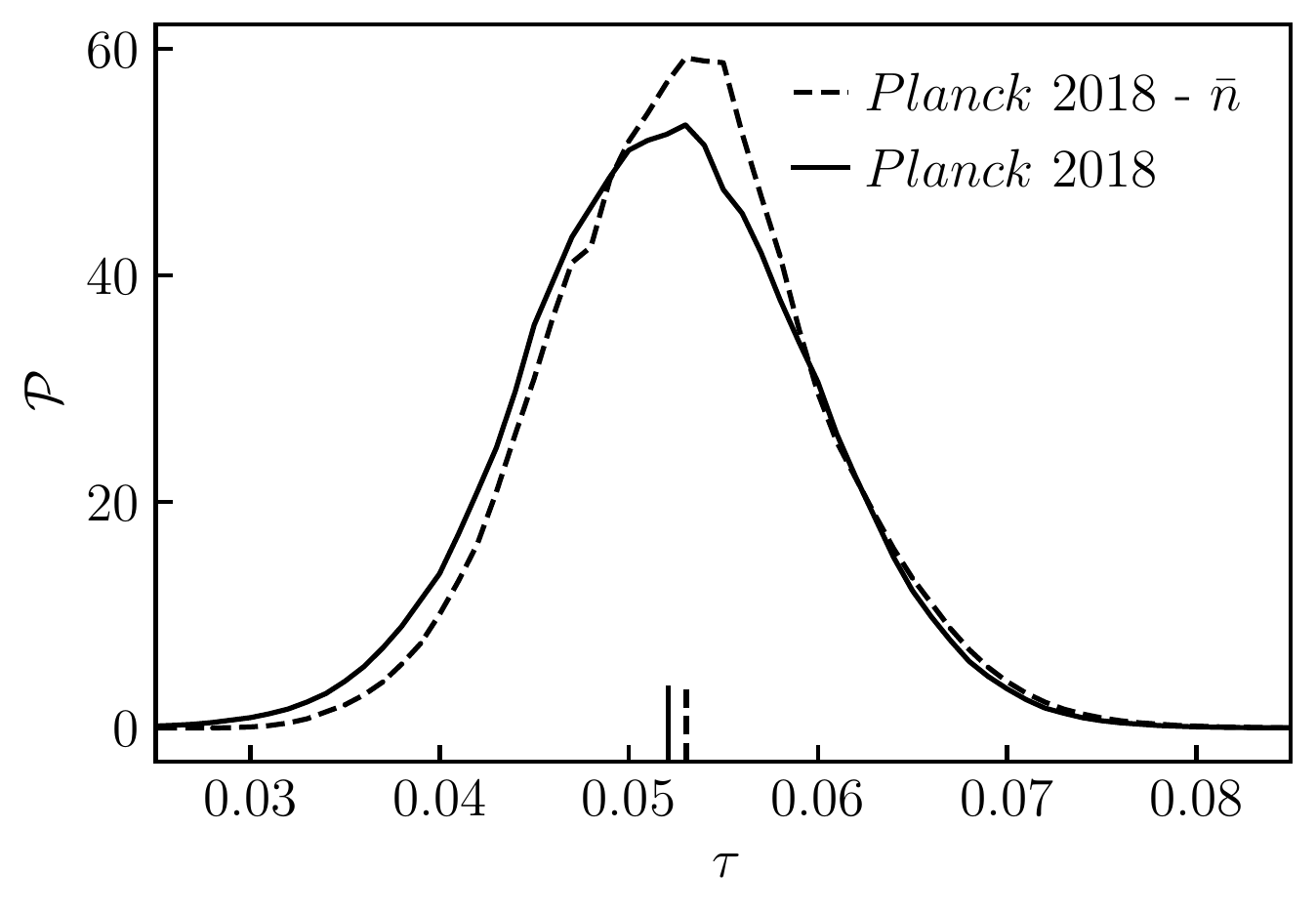}
    \vspace{-0.1in}
    \caption[tau posterior of Planck 2018 scaled to \SRone.]{Effect of smoothed noise template subtraction on the $\tau$ posterior computed from $100\times143$ full-mission cross-spectra of \Planck 2018 maps. This test is performed using the simulation-based likelihood \simlow.  The solid (dashed) line shows the posterior obtained when the subtraction is (is not) performed.}
    \label{fig:SRoll2_posterior}
    \vspace{-0.1in}
\end{figure}

This appendix discusses the effect of \emph{smoothed template} subtraction. The ADCNL effectively leads to CMB-independent offsets in \Planck HFI maps.  The CMB signal can then suffer chance correlations with these offsets, leading to additional scatter in power spectra.  This effect is seen in simulations and, as discussed in Sec.\ \ref{sec:NCM}, can potentially be mitigated by computing a smoothed estimate of the offset for each map and then subtracting the appropriate estimate from each of the QCS input maps.   Applying this prescription to   \SRone spectra leads to a  $\sim 10\%$ reduction in the posterior width for $\tau$ from \SRone and to a $\sim 5\%$ reduction from \SRtwo. This is consistent with the hypothesis that \SRtwo better reduces large-scale residuals than \SRone and so sees less of an improvement. Note that the procedure of removing smoothed templates from real data only leads one closer to the truth if the simulations from which the smoothed templates are computed are relatively accurate not only in the power spectrum domain but also in the map domain. 

Fig. \ref{fig:SRoll2_posterior} shows the effect of smoothed template subtraction on the $\tau$ posterior for \Planck 2018 data using \simlow ($EE$). The solid line yields a posterior of $\tau=0.0521 \pm 0.0077$ where no smoothed templates have been subtracted. The dashed line yields a posterior $\tau=0.0530 \pm 0.0071$.

\subsection{Comparison between \SRone and \SRtwo} 
\begin{figure}
    \centering
    \includegraphics[width=\columnwidth]{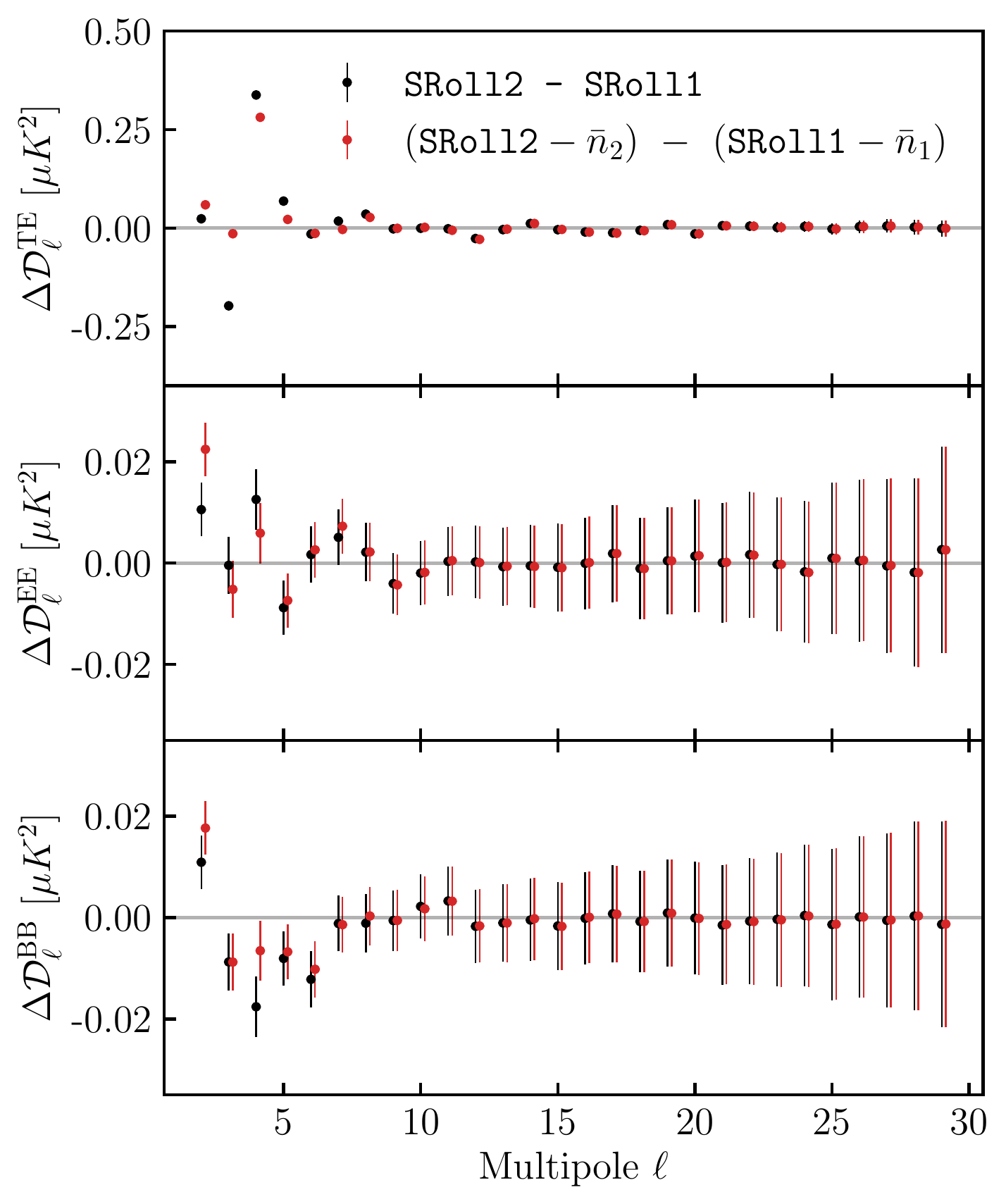}
    \vspace{-0.1in}
    \caption[Difference \SRone and \SRtwo.]{QCS estimates for $TE$, $EE$ and $BB$ $100\times143$ full-mission foreground-cleaned cross-spectra computed for the \emph{difference} between \SRtwo and \SRone maps.   The black points show the spectra computed before any smoothed templates are removed, while the red points show the spectra computed after the maps have their corresponding templates subtracted. The error bars for the cross-spectra are taken from the $BB$ spectrum as the signal has been removed. The $TT$ residual spectrum is not shown as it is consistent with zero. Note that the $BB$ errors plotted on the $TE$ spectrum are smaller than the points at low $\ell$; there are significant changes (relative to the $BB$ noise) between \SRone and \SRtwo in $TE$.}
    \label{fig:SRoll2-SRoll1}
    \vspace{-0.1in}
\end{figure}

Power spectra of the difference maps computed between \SRone and \SRtwo are shown in Fig. \ref{fig:SRoll2-SRoll1}. The error bars are based on the error estimates obtained from the $BB$ spectra since no signal should be left in the difference maps. Both maps have been foreground cleaned with the respective maps, as detailed in Sec. \ref{sec:CMB_data}. The red points show the spectra computed from maps that have corresponding smoothed templates (denoted $n_1$ for \SRone and $n_2$ for \SRtwo) subtracted. The black points show the difference at the map level without subtracting the templates. We note that the subtractions only have large effects for $\ell \leq 4$ (as one would expect given that this is the maximum multipole retained in the templates) and, except for the quadrupole, reduce the residuals. With or without the template subtractions, the residual cross-spectra show differences for the very low multipoles below $\ell\sim 10$ for $TE$, $EE$ and $BB$. We do not show the $TT$ spectrum since it is signal dominated and, as shown in Fig. \ref{fig:QCS_100x143_data}, the $TT$ spectra are almost identical for \SRone and \SRtwo. This test confirms that the \SRtwo map-making algorithm indeed alters large-scale polarisation features relative to \SRone. 

\section{Technical details for density-estimation likelihood-free inference} \label{sec:technical_details_delfi}

In the present Appendix, we will elaborate on the technical details of the likelihood-free inference code \pydelfi \citep{2019MNRAS.488.4440A}, based on theoretical ground work by \cite{2018PhRvD..98f3511L, 2018MNRAS.477.2874A, 2019MNRAS.488.4440A, 2018arXiv180509294L, 2018arXiv180507226P}. These methods provide a means of performing Bayesian inference when it is possible to produce high accuracy simulations of the observables in question, even when it is not possible to write down an explicit analytic expression for the likelihood. Examples of such problems include measuring cosmological parameters from type Ia supernovae \citep{2013ApJ...764..116W}, weak lensing peak counts \citep{2015A&A...583A..70L, 2020arXiv200908459J}, analysing the galaxy-halo connection \citep{2017MNRAS.469.2791H}, inferring photometric redshifts and size evolution of galaxies \citep{2017A&A...605A...9C}, measuring cosmological redshift distributions \citep{2018JCAP...02..042K}, estimating the ionising background from the Lyman-$\alpha$ and Lyman-$\beta$ forests \citep{2018ApJ...855..106D} and inferring the sum of the masses of Andromeda and the Milky Way \citep{2021PhRvD.103b3009L}.

The key principle behind LF methods is the understanding that a simulated observable, $\mat d$, and the input parameters, $\bf{\theta}$, are a sample from the joint probability distribution $\mathcal{P}(\mat d,\mathbf{\theta}|\mathcal{M})$ conditioned on the theoretical model. For compactness  we will hereafter suppress the conditional aspect of the distribution and make the assumption that our simulations accurately represent the true physical process\footnote{Note that a claim often made against LF methods is the explicit strong dependence on the accuracy of the simulations and that missing or inaccurate components in the simulations can bias inferences. Whilst this is the case, it is similarly true for other approaches that the validity of their result depends on the correctness of the chosen model.}. With a sufficiently large number of samples from  $\mathcal{P}(\mat d,\mathbf{\theta})$ we can obtain a posterior on the parameters by selecting the samples that have $\mat d_{\mathrm{data}}=\mat d_{\mathrm{sim}}$. In many situations, it is essentially impossible to obtain $\mat d_{\mathrm{data}}\equiv \mat d_{\mathrm{sim}}$ and thus $|\mat d_{\mathrm{data}}-\mat d_{\mathrm{sim}}| <\epsilon$ is used (for some suitable metric).

An immediate challenge to the above idea is that, for many situations including cases in cosmology, it can require an incredibly large number of simulations to obtain samples from the posterior, especially if one requires a small $\epsilon$ parameter. This is particularly true when the dimension of the observable is large. This challenge can be addressed with density estimation likelihood-free inference (\delfi). The conceptually simplest \delfi methods use a density estimator to obtain a parametric form for the joint distribution $\mathcal{P}(\mat d,\theta | w)$, where w are the parameters of the density estimator.  These methods have been found to require many orders of magnitude fewer simulations. In this work we use a slight variation: we use a density estimator to model the conditional distribution $\mathcal{P}(\mat d|\theta,w)$, utilising the fact that, when conditioned on the parameters, $\mat d_{\rm sim}$ is a sample from $\mathcal{P}(\mat d|\theta)$. This variation means that one can be agnostic about the properties of the chosen simulation points; see \cite{2019MNRAS.488.4440A} for a more detailed discussion. The likelihood is then obtained by evaluating this conditional distribution at the observed data $\mat d_0$. Given the likelihood and a prior, we then use MCMC sampling to obtain posterior samples. 

\subsection{Masked Autoregressive Flows}
We require the density estimation to be flexible enough to allow us to accurately approximate the true conditional distribution and also to be computationally tractable in order to both fit and use. Recent work by \eg \citet{NIPS2016_6084, 2016arXiv160502226U} have shown that Masked Autoregressive Flows (MAFs) provide one such method. 

MAFs rely on two steps to achieve these goals. Firstly they utilise the chain rule to express the multidimensional conditional distribution, $\mathcal{P}(\mat d|\mat \theta)$, as a series of one dimensional conditionals 
\begin{align} 
    \mathcal{P}(\mat d | \mat \theta) = \prod_{i=1}^{n} \mathcal{P}(d_i | \mat{d}_{1:i-1}, \mat{\theta})\ .
\end{align}
A model utilising this decomposition is known as an autoregressive model. Next, a form is chosen for the one dimensional conditional distributions. We assume the conditionals are one dimensional Gaussian distributions whose means, $\mat{\mu}$, and standard deviations, $\mat{\sigma}$, depend on  $\mat{d}_{1:i-1}$ and $\mat{\theta}$. To achieve the desired flexibility we use neural networks, with parameters $\mat{w}$, to parametrise these functions as $\mat{\mu}(\mat{d}_{1:i-1},\mat{\theta}|\mat{w})$ and $\mat{\sigma}(\mat{d}_{1:i-1},\mat{\theta}|\mat{w})$.  As was shown in \cite{2015arXiv150203509G} this setup can equivalently be formulated as developing a mapping from $\mat{d}$ to the variable $u_i = x_i- \mu_i(\mat{d}_{1:i-1},\mat{\theta}|\mat{w})/\sigma_i(\mat{d}_{1:i-1},\mat{\theta}|\mat{w})$ where $\mat{u}$ are independent zero mean, unit variance Gaussian random variables. This means the density estimator has the simple form
\begin{align} 
    \mathcal{P}(\mat d | \mat \theta, \mat w) =  \mathcal{N}[\mat u (\mat{d,\theta; w})|\mat{0, 1}] \times \prod_{i=1}^n \sigma_i^n(\mat{d, \theta; w}),
\end{align}
where the product over the standard deviations is the Jacobian from transforming from $\mat{u}$ to $\mat{d}$.
We use an efficient implementation of this setup called the Gaussian Masked Autoencoders for Density Estimation and hereafter refer to this method as MADE.

There are two main limitations for using MADE: first, they depend sensitively on the order of factorisation and, second, the assumption of Gaussian conditionals may be overly restrictive. To mitigate these shortcomings we stack a series of MADEs to make a MAF. The output $\mat u$ of each MADE is the input of the next one. The conditional density estimator is thus given by
\begin{align}
    \mathcal{P}(\mat d | \mat{\theta; w}) & = \prod_i \mathcal{P}(d_i | \mat{d}_{1:i-1}, \mat{\theta}) \nonumber \\
    &= \mathcal{N}[\mat u (\mat{d,\theta; w})|\mat{0, 1}] \times \prod_{n=1}^{N_{\rm MADE}} \prod_{i=1}^n \sigma_i^n(\mat{d, \theta; w}) \ .
\end{align}
Thus we obtain a conditional distribution that is both analytically tractable (a simple product of Gaussians) and highly flexible. 

Finally we fit the weights of the neural network as detailed in Sec. \ref{sec:Likelihood_DELFI} around Eq. \eqref{eq:KLdiv}. 
By minimising the negative loss function in Eq. \eqref{eq:loss_function} we train the neural density estimators (NDEs) with respect to the network weights. The problem of overfitting is mitigated by applying the standard machine learning procedures of early stopping, dropouts and the random selection of training and testing subsets. 

\subsection{Architecture of \pydelfi}
\begin{table}
\centering
\begin{tabular}{cccccc}
\hline\hline
No. & $N_{\rm MADE}$& \makecell{Hidden \\ Layers}   & Units & \makecell{Validation \\ Loss}& $\tau$ Posterior\\ \hline
1 & 20 &2 &50 & 26.0 & $0.0519 \pm 0.0071$ \\
2 & 10 &2 &50 & 25.9 &$0.0512 \pm 0.0071$\\
3 & 5 &2 &50  & 26.0 &$0.0526 \pm 0.0073$\\
4 & 3 &2 &75  & 26.0 &$0.0526 \pm 0.0069$\\
5 & 5 &2 &75  & 25.8 &$0.0529 \pm 0.0073$\\
6 & 10 &2 &75 & 25.9 &$0.0514 \pm 0.0070$\\
7 & 20 &2 &75 & 26.0 &$0.0519 \pm 0.0068$\\
8 & 5 &3 &35  & 26.0 &$0.0521 \pm 0.0067$\\\hline
\end{tabular}
\caption{Details of the eight MAFs making up the NDEs of the \pydelfi architecture used in this paper, along with the $\tau$ posteriors from each NDE for the \SRone $100 \times 143$ $EE$ analysis. (The individual posteriors are shown in Fig. \ref{fig:stacked_NDEs}.)}
\label{tab:architecture_NDE}
\vspace{-0.1in}
\end{table}

\begin{figure}
    \centering
    \includegraphics[width=\columnwidth]{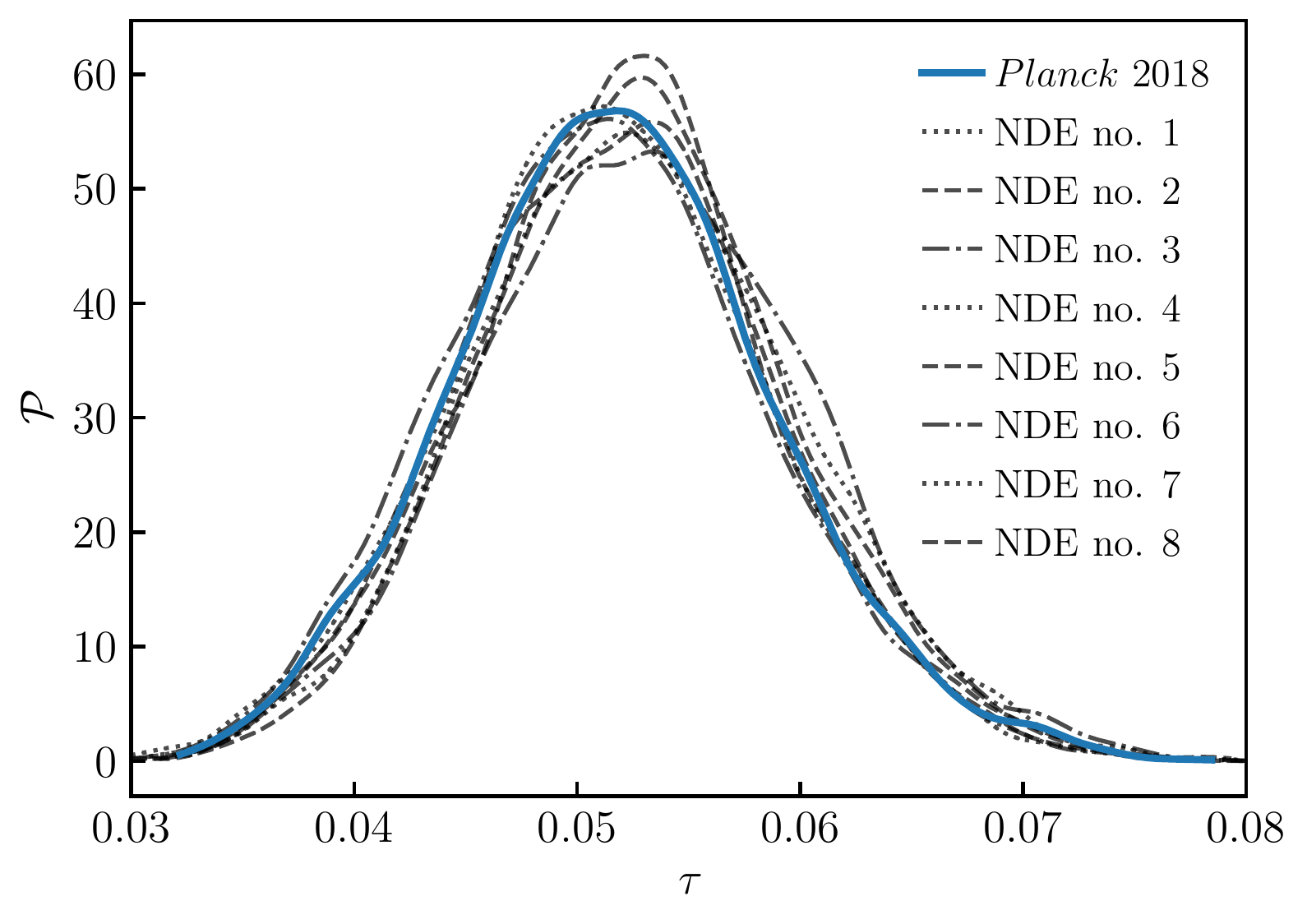}
    \vspace{-0.1in}
    \caption[stacked posteriors.]{\Planck 2018 posteriors obtained from our eight NDEs in our \pydelfi architecture.   The black lines show the individual constraints, and the blue solid line shows the weighted average that makes our \pydelfi result. (The means and standard deviations of the individual posteriors are given in Table \ref{tab:architecture_NDE}.)}
    \label{fig:stacked_NDEs}
    \vspace{-0.1in}
\end{figure}

For the density estimation, we select an ensemble of NDEs, all of them MAFs, with different numbers of MADEs, hidden layers and neurons per layer, as shown in Table \ref{tab:architecture_NDE}. All the NDEs use a $\tanh$ activation function, and are trained using the stochastic gradient optimiser \textsc{adam} \citep{2014arXiv1412.6980K}. To shorten training times of the NN, we trained the model on graphics processing units. To avoid overfitting, we use one tenth of the training set at each training cycle to validate our posterior (\ie perform `early-stopping') and use a learning rate of 0.001. 

The result from density estimation of the different NDEs is shown in Fig. \ref{fig:stacked_NDEs}. This is a powerful cross-check to determine whether or not our architecture was chosen appropriately, remembering that there is no general procedure for choosing NN architectures.  Each NDE yields a consistent posterior, also shown in Table \ref{tab:architecture_NDE}.  All the NDEs are then stacked and weighted by the loss evaluated during training,
\begin{equation}
    \mathcal{P}(\bf{t} | \bf{\theta; w}) = \sum_{\alpha = 1}^{N_{\rm NDE}}\beta_{\alpha}p_{\alpha}(\bf{t} | \bf{\theta; w})\ ,
\end{equation}
following \citep{2019MNRAS.488.4440A}, also illustrated in Fig.\ \ref{fig:stacked_NDEs}. This is motivated by the demonstration of \citet{NIPS1997_1353, Smyth_NDE} that stacking multiple NDEs is typically more robust than using an individual one.  (In fact, the high degree of consistency between the NDEs makes us suspect that using fewer of them would still be sufficient to constrain $\tau$, saving training time.) 

\subsection{Score and data compression} \label{sec:delfi_compression}

To compute a joint likelihood using low multipole temperature and polarisation QCS spectra ($TT$, $TE$, $EE$) an additional compression step is required to reduce the dimensionality of the problem. Therefore, the full $N$-dimensional data set $\mathbf{D}\in \mathbb{R}^N$ is first compressed to quadratic cross-spectra, \ie summary statistics $\mathbf{d} \in \mathbb{R}^M$, with $M < N$. Then, we score compress the vector of $M$ power spectrum measurements $\mathbf{d}$ into a vector of $n$ components $\mathbf{t}\in \mathbb{R}^n$, with $n<M<N$. In order to inform the choice of statistic for the score compression, an approximate form of the log-likelihood function $\mathcal{L}$ is assumed. The statistic $\mathbf{t}$ is then the gradient of the approximate log-likelihood, evaluated at some fiducial parameter values $\mathbf{\theta}_*$, i.e.\ $\mathbf{d} \mapsto \mathbf{t}=\nabla_{\mathbf{\theta}} \mathcal{L}_*$.

In the main body of the paper we compressed the $TT$, $TE$ and $EE$ cross-spectra separately for computational simplicity. This is suboptimal, as power spectrum elements between spectra are correlated. However this does not bias our cosmological constraints and \pydelfi does capture residual correlations between the individual $TT$, $TE$ and $EE$ statistics. The approximate log-likelihood we use for the score compression step is the analytic result that can be derived for cross-spectra computed on the full sky with Gaussian isotropic noise, namely the variance-gamma distribution \citep{10.2307/2332556}:
\begin{align} \label{eq:pearson}
    \mathcal{P}(\hat{C}^{XY}_{\ell}) &= \frac{(2\ell+1)\left| \hat{C}^{XY}_{\ell}\right|^{\ell}}{\Gamma(\frac{2\ell+1}{2})\sqrt{2^{2\ell}\pi (1-\rho_{\ell}^2)\left( C_{\ell}^{XX}C_{\ell}^{YY}\right)^{\ell+1}  } }  \nonumber \\ 
    & \quad \times K_{\ell}\left(\frac{(2\ell+1)\left|\hat{C}_{\ell}^{XY} \right| }{(1-\rho_{\ell}^2)\sqrt{ C_{\ell}^{XX}C_{\ell}^{YY}}} \right) \nonumber \\ 
    & \quad \times  \exp\left(\frac{ (2\ell+1)\rho_{\ell} \left|\hat{C}_{\ell}^{XY} \right|}{ (1-\rho_{\ell}^2) \sqrt{ C^{XX}_{\ell}C^{YY}_{\ell}}} \right)\ .  
\end{align}
Here $\Gamma(n)$ is the gamma function, $K_n(x)$ is the modified Bessel function of the second kind and $\rho_\ell \equiv C^{XY}/\sqrt{C^{XX} C^{YY}}$ is the correlation coefficient. Using a fiducial model with $\tau =0.06$, we differentiate Eq.\ \eqref{eq:pearson} to obtain
\begin{equation}
    t = \left.\nabla_{\mathbf{\tau}}\ln \mathcal{P}(\hat{C}^{XY}_\ell) \right\vert_{\theta_*},
\end{equation}
one each for $TT$, $TE$ and $EE$. 

We then fit the parameters of the \pydelfi likelihood using the compressed statistics of the simulations and the compressed data vector, yielding the pair $\{\tau, t \}$. For the case where we only consider $EE$, the compressed statistic is related to the maximum likelihood estimate for $\tau$; see \cite{2018MNRAS.477.2874A} for a more detailed discussion. The limitations of the compression steps are discussed further in \cite{2018MNRAS.477.2874A}.

\subsection{Effect of score compression on the maximum likelihood value of $\tau$ and comparison between \pydelfi and \momento}

\begin{figure}
    \centering
    \includegraphics[width=\columnwidth]{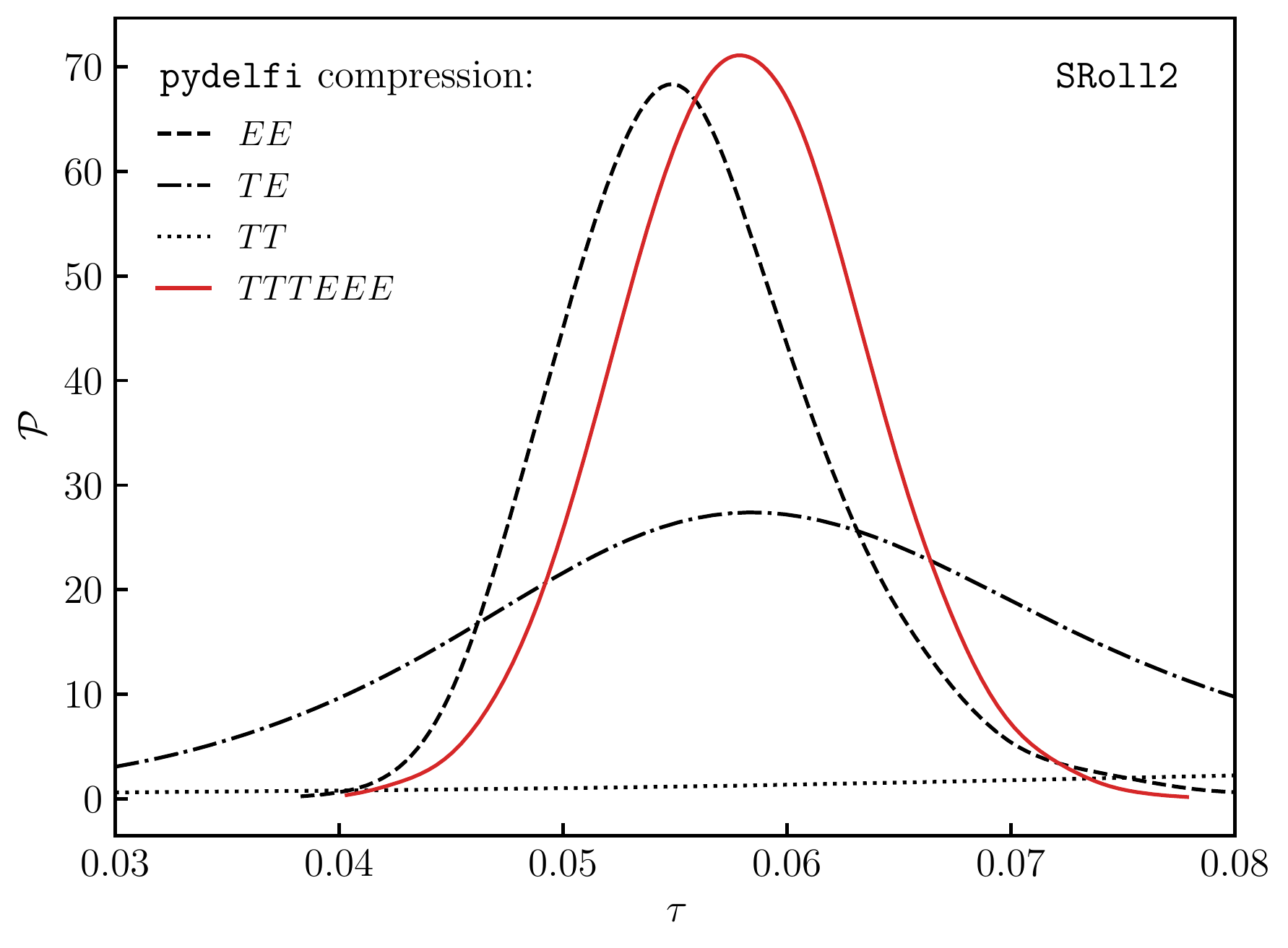}
    \vspace{-0.1in}
    \caption[\pydelfi score compression]{Score-compressed \pydelfi posteriors from \SRtwo $100\times143$ spectra. The individual $\tau$ posteriors for the $EE$ (dashed black line), $TE$ (dash-dotted black line) and $TT$ (dotted black line) statistics are shown, along with the
    result from the three-statistic $TTTEEE$ likelihood (solid red line).}
    \label{fig:compressed_posteriors}
    \vspace{-0.1in}
\end{figure}    

This section considers in more detail the properties of the score-compressed \pydelfi likelihoods, both amongst themselves and then in comparison with \momento. 

First, in Fig. \ref{fig:compressed_posteriors} we illustrate the score-compressed posteriors obtained with $100\times143$ spectra from \SRtwo frequency maps using \pydelfi. The different posteriors from the $EE$ (dashed line), $TE$ (dash-dotted line) and $TT$ (dotted line) statistics are shown, along with that from the combined three-statistic $TTTEEE$ likelihood (solid red line).  These all used the full multipole range ($2\leq \ell \leq 29$) for the construction of the statistic(s). The $EE$ posterior is most constraining, the $TE$ posterior somewhat so, whereas the $TT$ one hardly varies over the $\tau$ scan. In the three-statistic likelihood, the addition of $TE$ information moves the mean $\tau$ value upwards by $\sim 0.25\sigma$ from that from the $EE$-statistic likelihood.  This upwards shift is also seen for \momento in Table \ref{tab:short_result_summary} but is not apparent there for \pydelfi.  This is because the table shows results from the \emph{non-score-compressed} \pydelfi $EE$ likelihood and from the \emph{score-compressed} three-statistic \pydelfi $TTTEEE$ likelihood. The upwards shift from adding $TE$ partially cancels the downward movement caused in passing from a full $EE$ likelihood to a score-compressed one.

\begin{figure}
    \centering
    \includegraphics[width=\columnwidth]{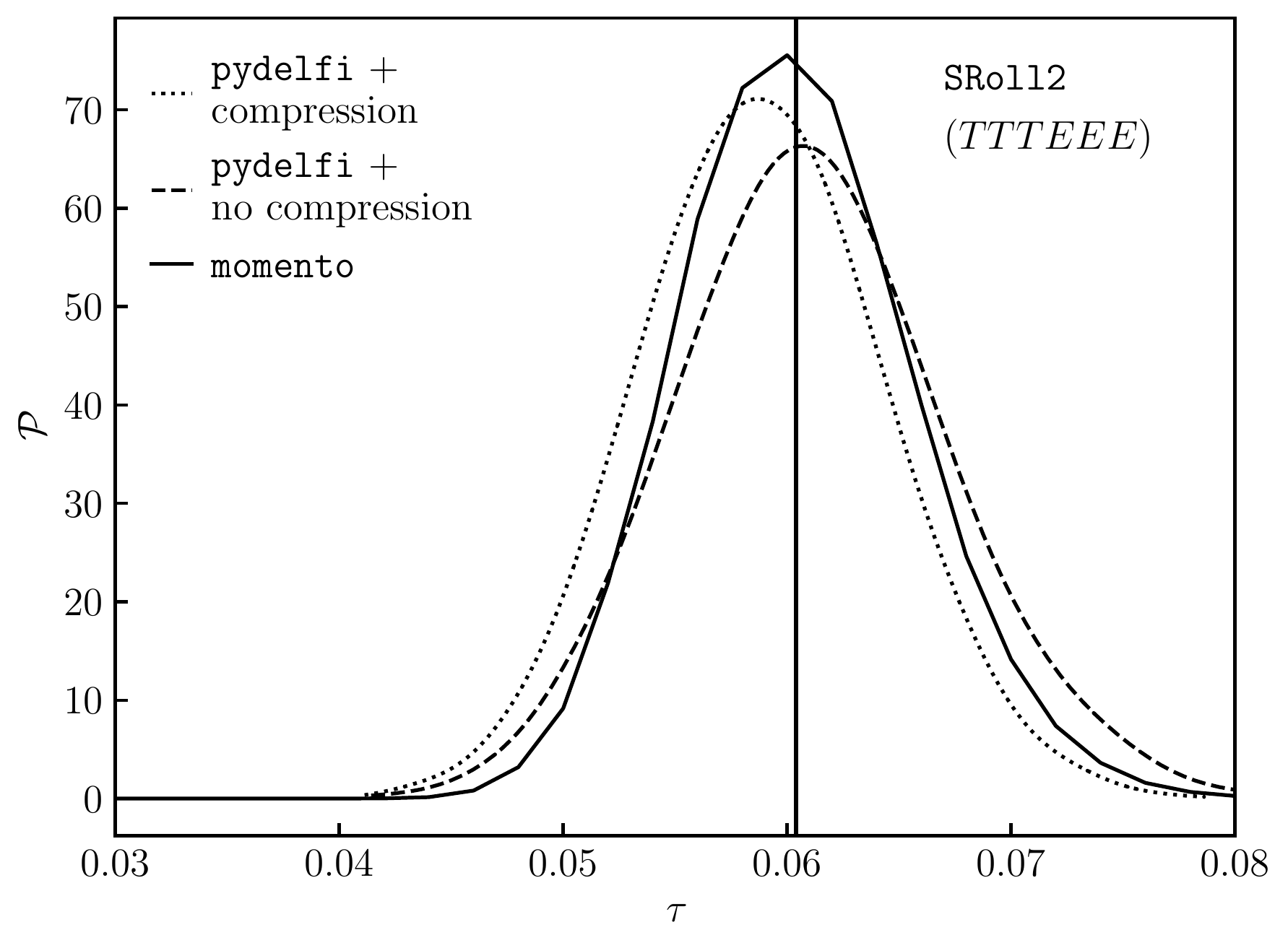}
    \vspace{-0.1in}
    \caption[\pydelfi score compression]{Comparison of $\tau$ posteriors from $100\times 143$ \SRtwo $TTTEEE$ spectra using \pydelfi with (dotted line) and without compression (dashed line) and using \momento (solid line). For this test all likelihoods use the same multipole range  of $2\leq \ell \leq 10$. We note that avoiding score compression for \pydelfi results in a shift upwards by $\sim 0.5 \sigma$ of the $\tau$ maximum likelihood value compared to that from the score compressed likelihood, leading to a similar result to \momento.}
    \label{fig:pydelfi_vs_momento_posterior_TTTEEE}
\end{figure}    

Second, to further understand how \pydelfi performs in comparison to \momento in the joint temperature-polarisation case, we produce a version of \pydelfi that does not need score compression.  By limiting the power spectrum multipoles to those that are most constraining for $\tau$, \ie $2\leq \ell \leq 10$, rather than $2\leq \ell \leq 29$, we have a computationally-manageable 27-dimensional problem, rather than a difficult 84-dimensional one. Comparing the posteriors with and without score compression on \SRtwo maps for the three-statistic $TTTEEE$ likelihoods, we obtain:
\begin{subequations}
\begin{align}
&\tau=0.0578\pm 0.0063,\quad ({\rm \pydelfi, \, \ell \leq 10 + comp.}), \label{eq:tau_compress1} \\
&\tau=0.0612\pm 0.0060,\quad ({\rm \pydelfi, \, \ell \leq 10 + no\ comp.}). \label{eq:tau_compress2}
\end{align}
\end{subequations}
Fig. \ref{fig:pydelfi_vs_momento_posterior_TTTEEE} illustrates these posteriors and compares them to the \momento posterior. This confirms the role of the score compression discussed above in affecting constraints on $\tau$.


\bsp	
\label{lastpage}
\end{document}